\documentclass[twocolumn]{aastex7}

\shorttitle{Photometry and SFHs for 36 UFDs}
\shortauthors{Durbin et al.}

\graphicspath{{./}{}}

\usepackage{amsmath,amssymb,xspace}
\usepackage{fp,mathtools}

\let\drizzlepac=\Drizzlepac

\newcommand{\TweakReg}{\texttt{TweakReg}\xspace}

\newcommand{\Astrodrizzle}{\texttt{AstroDrizzle}\xspace}

 \newcommand{\hstonepass}{\texttt{hst1pass}\xspace}
 \newcommand{\f}[1]{\mathrm{\uppercase{f{#1}}}}
 \let\F=\f

 \newcommand{\hst}{\emph{HST}\xspace}
 \let\HST=\hst

 \let\roman=\rst

 \newcommand{\gaia}{\emph{Gaia}\xspace}
 \let\Gaia=\gaia

\begin{document}

\title{The HST Legacy Archival Uniform Reduction of Local Group Imaging (LAURELIN). I. Photometry and Star Formation Histories for 36 Ultra-faint Dwarf Galaxies}

\correspondingauthor{Meredith Durbin}
\email{meredith.durbin@berkeley.edu}

\author[0000-0001-7531-9815]{Meredith J. Durbin}
\affiliation{Department of Astronomy, University of California, Berkeley, Berkeley, CA 94720, USA}
\email{meredith.durbin@berkeley.edu}

\author[0000-0003-1680-1884]{Yumi Choi}
\affiliation{NSF National Optical-Infrared Astronomy Research Laboratory, 950 N. Cherry Avenue, Tucson, AZ 85719 USA}
\email{yumi.choi@noirlab.edu}

\author[0000-0002-1445-4877]{Alessandro Savino}
\affiliation{Department of Astronomy, University of California, Berkeley, Berkeley, CA 94720, USA}
\email{asavino@berkeley.edu}

\author[0000-0002-6442-6030]{Daniel R. Weisz}
\affiliation{Department of Astronomy, University of California, Berkeley, Berkeley, CA 94720, USA}
\email{dan.weisz@berkeley.edu}

\author[0000-0002-6442-6030]{Andrew E. Dolphin}
\affiliation{Raytheon Technologies, 1151 E. Hermans Road, Tucson, AZ 85756, USA}
\affiliation{University of Arizona, Steward Observatory, 933 North Cherry Avenue, Tucson, AZ 85721, USA}
\email{adolphin@rtx.com}

\author[0000-0002-1264-2006]{Julianne J. Dalcanton}
\affiliation{Center for Computational Astrophysics, Flatiron Institute, 162 Fifth Avenue, New York, NY 10010, USA}
\affiliation{Department of Astronomy, University of Washington, Box 351580, Seattle, WA 98195-1580, USA}
\email{jdalcanton@flatironinstitute.org}

\author[0000-0001-6529-9777]{Myoungwon Jeon}
\affiliation{School of Space Research, Kyung Hee University, 1732 Deogyeong-daero, Giheung-gu, Yongin-si, Gyeonggi-do 17104, Republic of Korea}
\affiliation{Department of Astronomy \& Space Science, Kyung Hee University, 1732 Deogyeong-daero, Yongin-si, Gyeonggi-do 17104, Republic of Korea}
\email{myjeon@khu.ac.kr}

\author[0000-0002-3204-1742]{Nitya Kallivayalil}
\affiliation{Department of Astronomy, University of Virginia, 530 McCormick Road, Charlottesville, VA 22904, USA}
\email{njk3r@virginia.edu}

\author[0000-0002-9110-6163]{Ting S. Li}
\affiliation{Department of Astronomy and Astrophysics, University of Toronto, 50 St. George Street, Toronto ON, M5S 3H4, Canada}
\email{ting.li@astro.utoronto.ca}

\author[0000-0002-6021-8760]{Andrew B. Pace}
\affiliation{Department of Astronomy, University of Virginia, 530 McCormick Road, Charlottesville, VA 22904, USA}
\email{yyr6cm@virginia.edu}

\author[0000-0002-9820-1219]{Ekta~Patel}\thanks{NASA Hubble Fellow}
\affiliation{Department of Physics and Astronomy, University of Utah, 115 South 1400 East, Salt Lake City, Utah 84112, USA}
\email{ekta.patel@utah.edu}

\author[0000-0001-5618-0109]{Elena Sacchi}
\affiliation{Leibniz-Institut f\"{u}r Astrophysik Potsdam (AIP), An der Sternwarte 16, 14482 Potsdam, Germany}
\email{esacchi@aip.de}

\author[0000-0003-0605-8732]{Evan D. Skillman}
\affiliation{University of Minnesota, Minnesota Institute for Astrophysics, School of Physics and Astronomy, 116 Church Street, S.E., Minneapolis, MN 55455, USA}
\email{skill001@umn.edu}

\author[0000-0001-8368-0221]{Sangmo Tony Sohn}
\affiliation{Space Telescope Science Institute, 3700 San Martin Drive, Baltimore, MD 21218, USA}
\affiliation{Department of Astronomy \& Space Science, Kyung Hee University, 1732 Deogyeong-daero, Yongin-si, Gyeonggi-do 17104, Republic of Korea}
\email{tsohn@stsci.edu}

\author[0000-0001-7827-7825]{Roeland P. van der Marel}
\affiliation{Space Telescope Science Institute, 3700 San Martin Drive, Baltimore, MD 21218, USA}
\affiliation{Center for Astrophysical Sciences, The William H. Miller III Department of Physics \& Astronomy, Johns Hopkins University, Baltimore, MD 21218, USA}
\email{marel@stsci.edu}

\author[0000-0003-0603-8942]{Andrew Wetzel}
\affiliation{Department of Physics \& Astronomy, University of California, Davis, CA 95616, USA}
\email{awetzel@ucdavis.edu}

\author[0000-0002-7502-0597]{Benjamin F. Williams}
\affiliation{Department of Astronomy, University of Washington, Box 351580, Seattle, WA 98195-1580, USA}
\email{benw1@uw.edu}

\begin{abstract}
We present uniformly measured resolved stellar photometry and star formation histories (SFHs) for 36 nearby ($\lesssim 400$~kpc) ultra-faint dwarf galaxies (UFDs; $-7.1 \le M_V \le +0.0$) from new and archival \HST\ imaging.  We measure homogeneous distances to all systems via isochrone fitting and find good agreement ($\le 2\%$) for the 18 UFDs that have literature RR Lyrae distances.  From the ensemble of SFHs, we find: (i) an average quenching time (here defined as the lookback time by which 80\% of the stellar mass formed, $\tau_{80}$) of $12.48 \pm 0.18$ Gyr ago ($z = 4.6_{-0.5}^{+0.6}$), which is compatible with reionization-based quenching scenarios; and (ii) modest evidence of a delay ($\lesssim 800$~Myr) in quenching times of UFDs thought to be satellites of the LMC or on their first infall, relative to long-term Galactic satellites, which is consistent with previous findings.
We show that robust SFH measurement via the ancient main sequence turnoff (MSTO) requires a minimum effective luminosity (i.e., luminosity within the observed field of view) of $M_V \leq -2.5$, which corresponds to $\sim100$ stars around the MSTO. We also find that increasing the S/N above $\sim100$ at the MSTO does not improve SFH precision, which remains dominated by stochastic effects associated with the number of available stars. A main challenge driving the precision of UFD SFHs is limitations in the accuracy of foreground dust maps.
We make all photometry catalogs public as the first data release of a larger \HST\ archival program targeting all dwarf galaxies within $\sim1.3$~Mpc.
\end{abstract}


\section{Introduction}\label{sec:intro}
Deep photometric and spectroscopic campaigns over the two last decades have shown that ultra-faint dwarf galaxies \citep[UFDs; $M_V \ge -7.7$,][]{2019ARA&A..57..375S}, the least massive galaxies in the Universe with observable baryons, are extremely metal-poor, ancient, and dark matter dominated  systems \citep[e.g.,][]{2008AJ....135.1361D,2010ApJ...716..446S,2012ApJ...744...96O,2013ApJ...767..134V,2013ApJ...763...61G,2014ApJ...796...91B,2014ApJ...789..148W,2019ApJ...870...83J,2019ARA&A..57..375S,2021ApJ...920L..19S,2021ApJ...909..192G,2021ApJ...908...18S,2023ApJ...944...14M,2023ApJ...956...86S,2023ApJ...944...43S,2023ApJ...948...50W,2024ApJ...967..161M}. The observed properties of UFDs support theoretical ideas that galaxy growth in the lowest-mass halos is limited or entirely suppressed by \added{a combination of} supernova feedback and the ultraviolet (UV) background radiation resulting from cosmic reionization, which not only heats and expels halo gas, but also prevents the cooling and inflow of fresh gas from the surrounding intergalactic medium 
\citep[e.g.,][]{1992MNRAS.256P..43E,2000ApJ...539..517B,2002ApJ...572L..23S,2002MNRAS.333..177B,2003MNRAS.343..679B,2009ApJ...693.1859B,2009MNRAS.400.1593M,2009MNRAS.395L...6S,2010ApJ...710..408B,2010ApJ...708.1398T,2013MNRAS.432.1989S,2015MNRAS.453.1305W,2017ApJ...848...85J,2020MNRAS.494.2200K}. 
This limitation, driven by the interplay of internal and external feedback, in turn, is an important mechanism needed to reproduce galaxy counts at the low-mass regime within cold dark matter cosmological models \citep[e.g.,][]{2016MNRAS.456...85S,2016ApJ...827L..23W,2018MNRAS.478..548S,2019MNRAS.483.1314B,2019MNRAS.489.4574G,2021MNRAS.505..783F,2021MNRAS.507.4211E}. 

Cosmic reionization is thought to be a highly inhomogeneous process \citep[e.g.,][]{2006PhR...433..181F, 2014ApJ...793..113P, 2015MNRAS.447.3402B, 2015ApJ...813L..38D, 2017PhRvL.119b1301S, 2018PhRvD..98l3519F, 2024ApJ...976...93J}. 
Regions with the highest density were reionized first, while those with lower density, which could include field galaxies, the outskirts of the proto-Local Group or isolated proto-Large Magellanic Cloud (LMC)-mass groups, were ionized later, allowing prolonged star formation in low-mass halos \citep[e.g.,][]{2004ApJ...613....1F, 2007MNRAS.377.1043M, 2010Natur.468...49R, 2013MNRAS.432L..51S, 2021MNRAS.502.6044E, 2025arXiv250302927Z}. 
Various simulations suggest that the expected reionization time difference among progenitors of MW or M31-size galaxies is expected to be 200-500~Myr \citep[e.g.,][]{2012ApJ...746..109L, 2014ApJ...785..134L,2018ApJ...856L..22A,2018MNRAS.480.1740D,2019ApJ...882..152Z, 2020MNRAS.494.2200K, 2020MNRAS.496.4087O, 2022MNRAS.515.2970S, 2023ApJ...959...31K,2025arXiv250716245Z}. 
\added{However, the impact of patchy reionization on low-mass galaxies, particularly in the context of surviving satellites of individual MW-mass halos, has not been studied as thoroughly due to computational limitations.}

Unfortunately, our ability to conduct detailed, empirical characterizations of the impact of reionization on the formation of UFDs remains limited.  
UFDs are generally too faint to be directly detected beyond the very local Universe, and our ability to place their formation history into a cosmological context is largely limited to the Local Group.  
Within the Local Group, detailed color-magnitude diagram (CMD)-based star formation histories (SFHs) have the potential to reveal early Universe signatures, such as patchy reionization. 
However, to date they have been restricted to small samples, primarily in the halo of the Milky Way \citep[MW, e.g.,][]{2012ApJ...748...88W, 2014ApJ...789..148W, 2014ApJ...796...91B,2021ApJ...920L..19S,2023ApJ...944...43S,2024arXiv240719534G}. 

It is only recently that studies have begun to explore how the duration and decline of star formation in UFDs varies across stellar mass and galaxy environment \citep{2021ApJ...920L..19S,2023ApJ...944...14M,2023ApJ...956...86S}.
Furthermore, precise proper motions (PMs) measured by the \Gaia\ mission have provided a novel opportunity to probe the environmental dependence of UFD's early evolution. By using Gaia PM-based orbital histories, the MW UFDs have been classified into 3 kinematic groups \citep[e.g.,][]{2018A&A...619A.103F,2018ApJ...867...19K,2020ApJ...893..121P,2022ApJ...940..136P}. 
The first is made of UFDs that are likely on their first infall into the MW’s halo. 
The second contains UFDs consistent with being long-term MW satellites. 
The third consists of UFDs associated with the LMC group.  The LMC is thought to be on its first approach to the MW \citep[e.g.,][]{2007ApJ...668..949B, 2013ApJ...764..161K}, bringing its satellites---the Small Magellanic Cloud (SMC) and potentially a number of UFDs---with it. 
Gaia PMs indicate that 4 of the known UFDs (Carina~I, Carina~II, Horologium~I, and Hydrus~I) are highly likely satellites of the LMC, and Phoenix~II and Reticulum~II are also suspected companions based on their position, distance, and orbital properties \citep{2017MNRAS.465.1879S, 2018ApJ...867...19K,2020ApJ...893..121P}.

While we acknowledge the complexity in reconstructing the early environments of surviving satellites \citep[e.g.,][]{2015ApJ...807...49W,2023MNRAS.518.1427S}, the different categories of UFD orbital histories in the MW halo mean that their ancient SFHs may carry the imprint of the different environmental conditions they might have experienced early in their evolution. Thus, despite the limitations of current information, examining the kinematic group context of UFDs can still provide valuable insights into possible environmental influences on their early evolutionary pathways.
For instance, by analyzing deep \hst\ imaging, \citet{2021ApJ...920L..19S} measured a difference of $\sim 600$~Myr between the average quenching time of 3 UFDs associated with the LMC and those of 4 other UFDs. 
This work demonstrates how one might probe the inhomogeneity of reionization over co-moving volumes of $\sim350 \, \rm Mpc^3$ \citep{2016MNRAS.462L..51B} by quantifying the differences in SFH and quenching among MW UFDs, and by identifying their early environments by means of orbital reconstruction.

A robust characterization of these subtle differences in such ancient stellar populations requires careful mitigation of both statistical and systematic uncertainties. 
The sparse stellar populations of these low-mass galaxies can result in statistically uncertain quenching times on a per-object basis \citep{2014ApJ...796...91B}. 
On the other hand, systematic inhomogeneities in photometric reduction, stellar models, and/or distance determination that can stem from collating literature results can easily confound the signal encoded in the SFHs. 
The best approach is therefore to analyze a large sample of galaxies using a homogeneous methodology.

In this paper, we uniformly reduce and analyze all public deep \HST ACS/WFC optical imaging that has targeted the oldest main sequence turn-off (oMSTO) of 36 UFDs in the Local Group. 
We produce photometric catalogs, derive homogeneous distances, and measure the SFHs using consistent methodology. 
We use this dataset to analyze the formation and quenching of UFDs in the early Universe, with an emphasis on exploring putative differences within the UFD population as a function of kinematical association.

While there have been efforts toward the homogeneous analysis of deep \hst\ imaging in the Local Group \citep[e.g.,][]{2007IAUS..241..290G, 2014ApJ...796...91B, 2014ApJ...789..147W, 2021ApJ...920L..19S, 2025ApJ...979..205S}, most dwarf galaxies within $\sim 1.5$ Mpc have been the subject of small-sample studies, often with meaningful methodological differences between them, and/or include data of heterogeneous quality, which can introduce additional systematics \citep[e.g.,][]{2011ApJ...743....8W, 2021ApJ...909..192G, 2021ApJ...908...18S, 2023ApJ...944...43S, 2023ApJ...948...50W}.
In this paper, we make a step forward by providing a homogeneous photometric dataset that covers all MW UFDs with contemporary, deep \hst\ data. 
In future papers, we will expand our work by re-analyzing a larger number of galaxies, increasing the diversity of masses, morphologies, and environments in our database using a uniform approach across the entire sample. 
The high-level science products (HLSPs) produced by this program are hosted on MAST and can be accessed at the URL \url{https://archive.stsci.edu/hlsp/laurelin/} \added{or by \dataset[DOI: 10.17909/b8yw-wv58]{https://doi.org/10.17909/b8yw-wv58}.}

This paper is organized as follows. 
In \autoref{sec:obs} we provide an overview of our dataset and of the photometric reduction.
In \autoref{sec:SFHs} we detail our procedure for measuring the SFHs and quantifying the star formation quenching timescales. 
In \autoref{sec:discussion} we compare our results to past empirical studies and theoretical predictions, and make recommendations for future work. 
In \autoref{sec:conc} we present our conclusions.

\section{Observations and Data Reduction} \label{sec:obs}

\subsection{Sample}

\begin{figure}
    \centering
    \includegraphics[width=\linewidth]{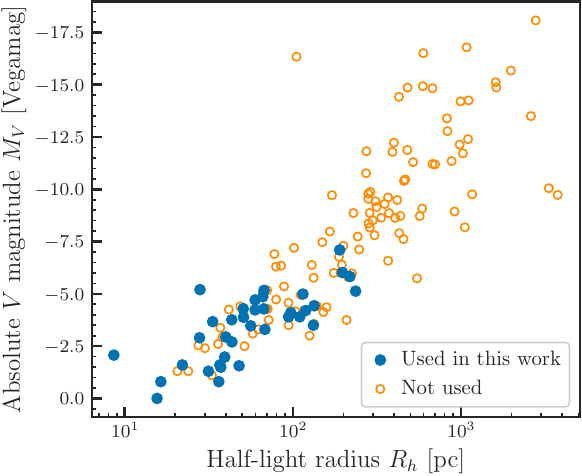}
    \caption{Size-magnitude relation for known Local Group dwarf galaxies within 1.5 Mpc. Filled blue points are the galaxies used in this work, and open orange circles are other local galaxies not studied here, either because they are too massive to be considered UFDs or because they lack \HST imaging. Data obtained from the Local Volume Database \citep{2024arXiv241107424P}.}
    \label{fig:size_mag}
\end{figure}

\begin{table*}[ht]
    \centering
    \caption{Observations} \label{tab:observations}
\begin{tabular}{llrrrrRRRrl}
\hline \hline
Galaxy & Abbrev. & \multicolumn{2}{c}{$t_{\rm exp}$ [ks]} & \multicolumn{2}{c}{$m_{50\%}$ [Vegamag]} & \rm{S/N}_{\rm MSTO} & f_{\star}^{a} & M_{V\rm{eff}}^{b} & Tiles & Proposal IDs \\
 & & F606W & F814W & F606W & F814W & & & & \\
\hline
Bo{\"o}tes I & Boo I & 41.1 & 41.2 & 27.58 & 26.94 & 335 & 0.14 & -3.89 & 5 & 12549, 15317 \\
Bo{\"o}tes II & Boo II & 9.2 & 9.2 & 27.43 & 26.76 & 427 & 0.35 & -1.81 & 2 & 14734 \\
Canes Venatici II & CVn II & 31.6 & 20.9 & 28.46 & 27.84 & 269 & 0.63 & -4.66 & 1 & 12549, 14236 \\
Cetus II & Cet II & 4.6 & 4.6 & 27.35 & 26.74 & 683 & 0.50 & 0.75 & 1 & 14734 \\
Columba I & Col I & 4.7 & 4.7 & 27.47 & 26.83 & 81 & 0.39 & -3.18 & 1 & 14734 \\
Coma Berenices & ComBer & 98.6 & 65.8 & 27.45 & 26.82 & 493 & 0.58 & -3.68 & 12 & 13449, 12549 \\
Draco II & Dra II & 4.7 & 4.7 & 27.39 & 26.74 & 874 & 0.29 & 0.53 & 1 & 14734 \\
Eridanus II & Eri II & 12.8 & 28.6 & 28.04 & 27.42 & 55 & 0.55 & -6.45 & 1 & 14234, 14224 \\
Eridanus III & Eri III & 4.6 & 4.6 & 27.41 & 26.74 & 198 & 0.96 & -2.03 & 1 & 14734 \\
Grus I & Gru I & 4.8 & 4.8 & 27.48 & 26.84 & 139 & 0.59 & -2.90 & 1 & 14734 \\
Grus II & Gru II & 4.6 & 4.6 & 27.43 & 26.81 & 344 & 0.10 & -1.45 & 1 & 14734 \\
Hercules & Her & 69.3 & 60.0 & 28.18 & 27.51 & 234 & 0.23 & -4.23 & 4 & 12549, 14236, 15182 \\
Horologium I & Hor I & 4.6 & 4.6 & 27.48 & 26.83 & 218 & 0.55 & -3.11 & 1 & 14734 \\
Horologium II & Hor II & 9.3 & 9.3 & 27.53 & 26.92 & 235 & 0.58 & -0.97 & 2 & 14734 \\
Hydra II & Hya II & 4.7 & 5.8 & 27.37 & 26.76 & 114 & 0.58 & -4.27 & 1 & 14734, 14224 \\
Hydrus I & Hyi I & 2.2 & 2.6 & 26.51 & 25.91 & 258 & 0.14 & -2.54 & 2 & 16293 \\
Leo IV & Leo IV & 30.9 & 20.5 & 28.28 & 27.66 & 253 & 0.25 & -3.49 & 1 & 12549, 14236 \\
Leo V & Leo V & 4.6 & 4.6 & 27.21 & 26.58 & 86 & 0.75 & -3.98 & 1 & 14770 \\
Pegasus III & Peg III & 4.7 & 4.7 & 27.25 & 26.58 & 68 & 0.47 & -3.29 & 1 & 14734 \\
Phoenix II & Phe II & 4.6 & 4.6 & 27.41 & 26.76 & 195 & 0.53 & -2.01 & 1 & 14734 \\
Pictor I & Pic I & 4.8 & 4.8 & 27.47 & 26.80 & 139 & 0.80 & -3.43 & 1 & 14734 \\
Pisces II & Pisc II & 4.7 & 4.7 & 27.39 & 26.74 & 88 & 0.71 & -3.87 & 1 & 14734 \\
Reticulum II & Ret II & 16.4 & 18.3 & 26.77 & 26.11 & 346 & 0.59 & -3.30 & 12 & 14734, 14766 \\
Reticulum III & Ret III & 4.7 & 4.7 & 27.38 & 26.72 & 178 & 0.42 & -2.36 & 1 & 14734 \\
Sagittarius II & Sag II & 4.6 & 4.6 & 27.20 & 26.58 & 272 & 0.59 & -4.63 & 1 & 14734 \\
Segue 1 & Seg 1 & 9.2 & 9.2 & 27.35 & 26.71 & 621 & 0.30 & -0.01 & 2 & 14734 \\
Segue 2 & Seg 2 & 4.6 & 4.6 & 27.37 & 26.74 & 430 & 0.20 & -0.24 & 1 & 14734 \\
Triangulum II & Tri II & 9.2 & 9.2 & 27.31 & 26.67 & 576 & 0.47 & -0.79 & 2 & 14734 \\
Tucana II & Tuc II & 17.5 & 18.6 & 27.44 & 26.79 & 332 & 0.22 & -2.25 & 4 & 14734 \\
Tucana III & Tuc III & 9.3 & 9.3 & 27.42 & 26.76 & 777 & 0.19 & 0.29 & 2 & 14734 \\
Tucana IV & Tuc IV & 9.3 & 9.3 & 27.48 & 26.85 & 419 & 0.08 & -0.79 & 2 & 14734 \\
Tucana V & Tuc V & 4.7 & 4.7 & 27.46 & 26.84 & 339 & 0.43 & -0.67 & 1 & 14734 \\
Ursa Major I & UMa I & 45.9 & 33.5 & 27.56 & 26.92 & 184 & 0.31 & -3.86 & 9 & 12549, 14236 \\
Ursa Major II & UMa II & 9.3 & 9.3 & 27.45 & 26.78 & 523 & 0.05 & -1.11 & 2 & 14734 \\
Virgo I & Vir I & 2.4 & 2.3 & 26.97 & 26.33 & 145 & 0.59 & -0.23 & 1 & 15332 \\
Willman 1 & Wil 1 & 4.6 & 4.6 & 27.42 & 26.78 & 427 & 0.35 & -1.77 & 1 & 14734 \\
\hline
\multicolumn{11}{l}{$^a$ The fraction of the galaxy's integrated light within the \HST footprint, calculated using the profiles shown in \autoref{fig:footprints}.} \\
\multicolumn{11}{l}{$^b$ The effective integrated $V$ magnitude of the \HST pointing, $M_{V\rm{eff}} = M_V - 2.5\log_{10}(f_\star)$.}
\end{tabular}
\end{table*}

\begin{figure*}
    \centering
    \includegraphics[width=\linewidth]{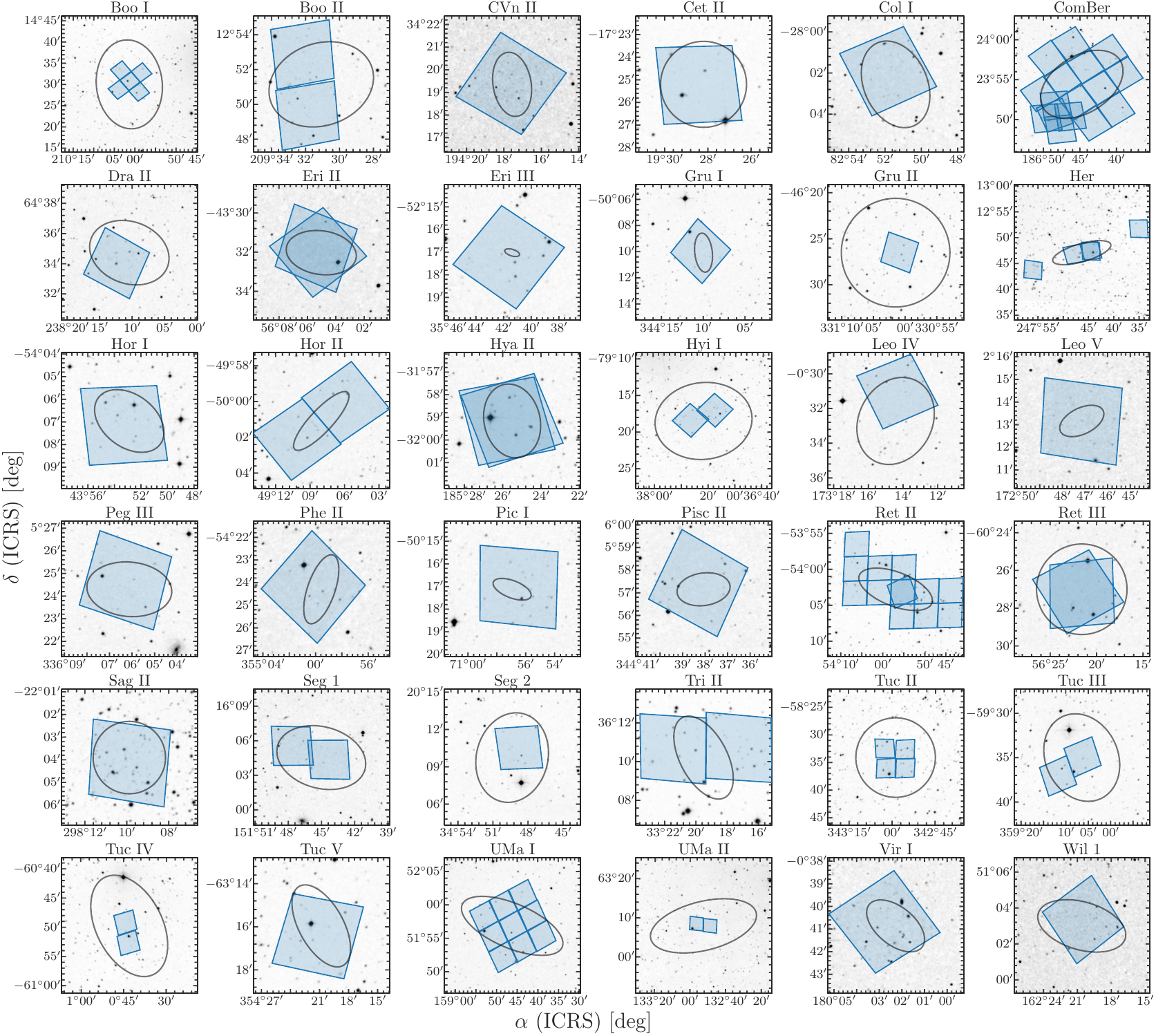}
    \caption{Footprints of all \hst observations used in this work (blue filled patches) overlaid on DSS2 imaging cutouts.
    Open black ellipses show the galaxy profiles at one half-light radius.
    }
    \label{fig:footprints}
\end{figure*}

Our sample comprises 36 UFDs with deep \HST ACS/WFC imaging in F606W and F814W (S/N $>$ 50 at the oMSTO in all cases, S/N $>$ 100 in most), spanning an absolute $V$ magnitude range of of $-7.1 \leq M_V \leq +0.0$.
The size-magnitude parameter space of our sample is shown in \autoref{fig:size_mag}.

We note that while most of the UFDs in this analysis have been confirmed spectroscopically or are too large to be globular clusters, there are three systems whose classifications remain uncertain: Eri III, Sag~II, and Tuc III \citep{2021MNRAS.503.2754L, 2023ApJ...958..167F, 2024ApJ...976..256S, 2025arXiv250305927Z}.
Of these, Sag~II is the most likely to be a cluster based on its very low mass-to-light ratio and minimal metallicity dispersion \citep{2021MNRAS.503.2754L}.
We retain these objects in our overall sample and in calculations that include the full sample, and refer to them as UFDs throughout regardless, 
but exclude Sag~II from certain parts of the later analysis that are restricted to a high-quality subsample.

Our full set of new and archival observations are summarized in \autoref{tab:observations}, and footprints of all pointings overlaid with galaxy profiles are shown in \autoref{fig:footprints}. 
The new data comprises two pointings in Hydrus~I (GO-16293, PI Choi),\footnote{GO-16293 also observed three fields in Carina~II, but the central pointing experienced a guide star failure and was not reobserved. We elected not to include the remaining two fields in this work, as we found they were too sparse to be of use for the goals of this paper.} and the bulk of the archival data comes from programs GO-14734 (PI Kallivayalil, 26 galaxies) and GO-12549 (PI Brown, 6 galaxies).
Other archival programs include GO-13449 (PI Geha), GO-14224 (PI Gallart), GO-14234 (PI Simon), GO-14236 (PI Sohn), GO-14766 (PI Simon), GO-14770 (PI Sohn), GO-15182 (PI Sand), GO-15317 (PI Platais), and GO-15332 (PI Crnojevic).

Following \citet{2025ApJ...979..205S} we quantify the fraction of a galaxy's total light expected to fall within the \HST pointing(s) as $f_\star$, which we estimate using surface brightness profiles.
From this and $M_V$ we calculate $M_{V\rm{eff}}$, the effective magnitude of the \HST observations, $M_V - 2.5\log_{10}(f_\star)$.
We adopt structural parameters from the exponential profile fitting results of \citet{2018ApJ...860...66M} where available, with the following exceptions: Cet~II \& Ret~III \citep{2015ApJ...813..109D}; Col~I \citep{2017AJ....154..267C}; Dra~II \citep{2015ApJ...813...44L}; Gru~II, Tuc~IV, \& Tuc~V \citep{2020ApJ...892..137S}; Hyi~I \citep{2018MNRAS.479.5343K}; Peg~III \citep{2022ApJ...933..217R}; Sag~II \& Tuc~III \citep{2018ApJ...863...25M}; Tuc~II \citep{2015ApJ...807...50B}; and Vir~I \citep{2016ApJ...832...21H}.

\subsection{Astrometric alignment} \label{ssec:hst_alignment}

We adopt a modified version of the workflows described in \citet{2017wfc..rept...19B}\footnote{\url{https://github.com/spacetelescope/gaia\_alignment/}} and \citet{2020AAS...23510907B}\footnote{\url{https://github.com/spacetelescope/wfc3\_photometry/}} to align our \HST exposures with \TweakReg.
\TweakReg is a routine included in the \drizzlepac software package \citep{2012drzp.book.....G, 2012ascl.soft12011S, 2015ASPC..495..281A, 2021AAS...23821602H} that derives transformations between coordinates of sources in a reference frame and sources in individual exposures using a triangle pattern matching algorithm, and optionally updates the exposure headers' WCS information with the ``tweaked" astrometric solutions.
Although absolute alignment to the \gaia DR3 reference frame is preferred for sufficiently populated images, a number of observations have few or no \gaia sources within their footprints.
Therefore we perform only relative alignment on a per-target basis.

We measured input exposure-level catalogs with the \hstonepass{} software \citep{2022wfc..rept....5A} for use as \TweakReg inputs.
\hstonepass{} is a publicly available FORTRAN routine designed to measure precise astrometry and photometry of relatively bright, isolated stars in individual \hst \texttt{flt} or \texttt{flc} frames using effective point spread function fitting \citep[ePSF,][]{2000PASP..112.1360A, 2006acs..rept....1A}.
With these we fit shifts, rotation, and scale relative to the deepest exposure for all exposures in a stack. The median RMS of the residuals is under 0.05 pixels, or 2.5 mas.
We then combined the aligned images to create a deep undistorted reference image for each target with \Astrodrizzle \citep{2002PASP..114..144F} with the parameters recommended for ACS in the Hubble Advanced Products pipeline \citep[HAP,][]{2022acs..rept....3M}.

We provide all drizzled reference images as part of our data release.

\subsection{Photometry}

\begin{figure*}
    \centering
    \includegraphics[width=\linewidth]{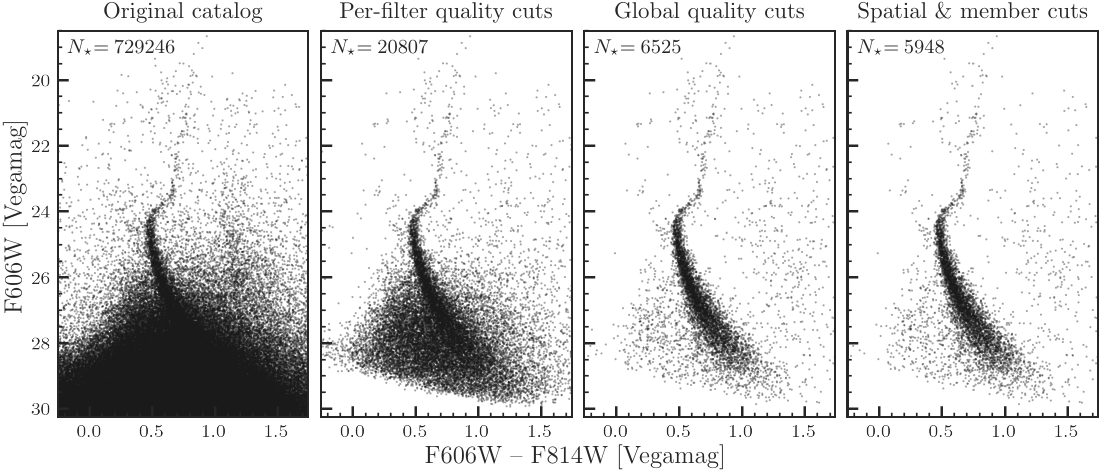}
    \caption{Application of successive culling criteria to CMDs of Hercules, with the fourth panel showing the final version of the catalog used in the remainder of this work. The stellar population signal of the galaxy remains constant throughout, whereas noise, artifacts, and contaminant populations are dramatically reduced.}
    \label{fig:cuts}
\end{figure*}

We measured full-stack photometry on the aligned images using the DOLPHOT package \citep{2000PASP..112.1383D, 2016ascl.soft08013D}.
We largely follow the input parameter recommendations of the PHAT \citep{2014ApJS..215....9W, 2023ApJS..268...48W} and PHATTER \citep{2021ApJS..253...53W} surveys, with the following exceptions:
\begin{enumerate}
    \item We turn DOLPHOT's internal alignment step off with \texttt{Align\,=\,0}. 
    In combination with \texttt{UseWCS\,=\,2}, this ensures that the \hstonepass-derived astrometric solutions described in the previous section are used as-is.
    \item Following \citet{2023ApJ...956...86S}, we add a second iteration of PSF fitting with \texttt{PSFPhotIT\,=\,2}, which improves noise estimates in the final photometric solution.
    (Note that this is distinct from the number of star-finding iterations, which is set by \texttt{SecondPass}.)
\end{enumerate}

We adopt somewhat stricter quality cuts on our photometry than the GST (``good star") cuts applied to PHAT \& PHATTER, as our data have much lower crowding. These cuts, applied per filter, are:
\begin{itemize}
    \item S/N $\geq$ 4
    \item Sharpness$^2$ $\leq$ 0.2
    \item Crowding $\leq$ 0.75
    \item Roundness $\leq$ 0.3
    \item Flag $\leq$ 3
\end{itemize}

\begin{figure*}
    \centering
    \includegraphics[width=0.98\textwidth]{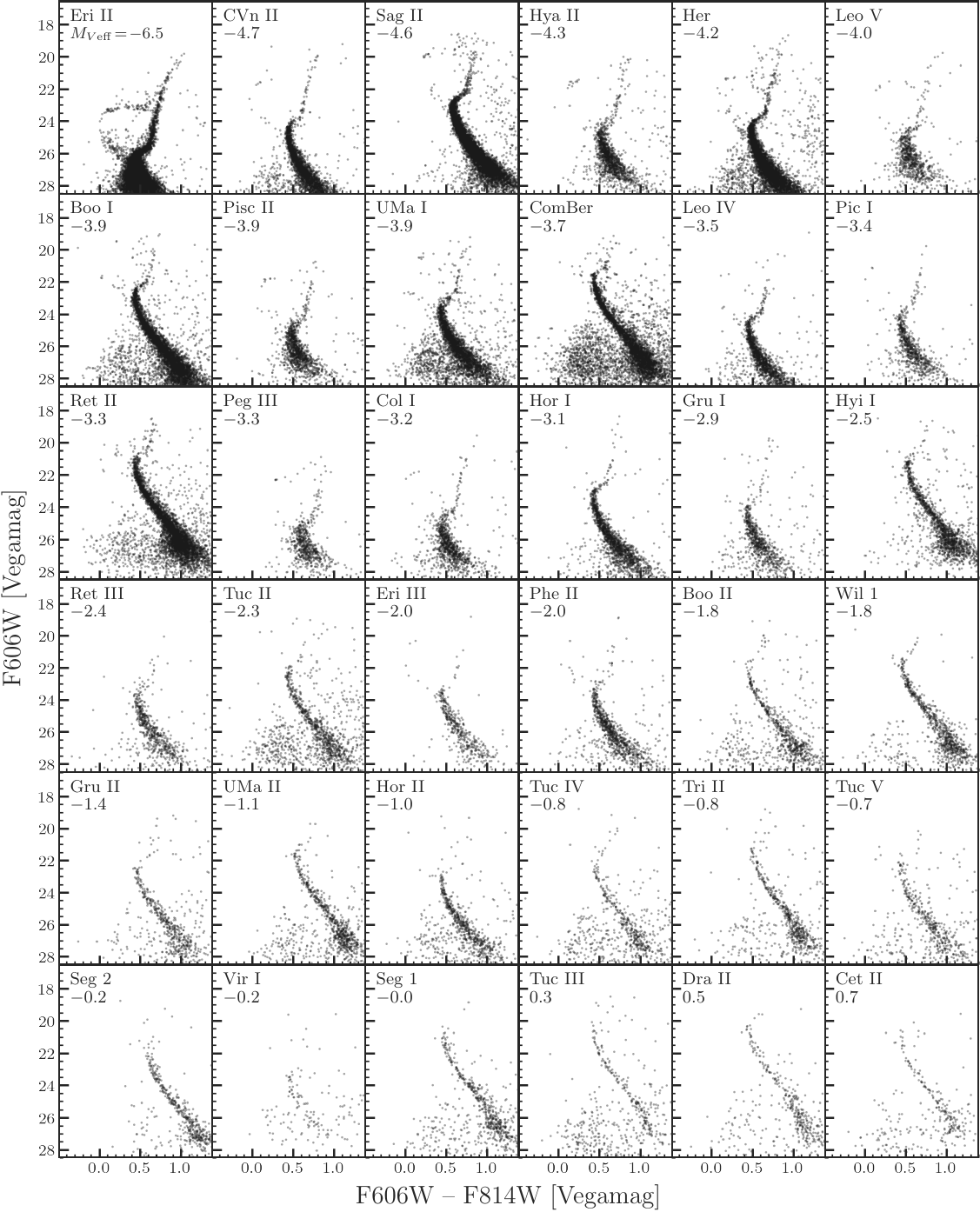}
    \caption{Color-magnitude diagrams of all galaxies, with all photometric quality and membership cuts applied.  The integrated magnitudes in each panel reflect the effective luminosity, i.e., the luminosity within \hst's field of view for each galaxy.}
    \label{fig:cmds}
\end{figure*}

We then impose further quality criteria on the sharpness, roundness, and crowding values by combining the filter-level quantities. 
We add the per-filter sharpness and roundness values in quadrature, and convert the crowding values from magnitude to fractional flux before taking their average and converting back to magnitudes to arrive at global values per source.
We apply a modified version of the culling technique explored by \citet{2015ApJ...814....3D}, who rejected low-quality measurements by fitting ellipses to two-dimensional distributions of the aforementioned quality metrics.
In our case, we require:
\begin{itemize}
    \item Sharpness$^2$ + Roundness$^2$ + Crowding$^2$ $<$ 0.1
    \item $|$Sharpness $\times$ Roundness $\times$ Crowding$|$ $<$ 0.001
\end{itemize}
In our uncrowded, high S/N data, these fairly strict cuts produce CMDs with substantially reduced populations of background galaxies and artifacts such as shredded diffraction spikes of saturated stars, while maintaining high completeness for the bona-fide stellar sources, as demonstrated in \autoref{fig:cuts} and \autoref{fig:cmds}.
Some galaxies, most notably Her, Boo~I, UMa~I, ComBer, and Ret~II, still show residual populations of unrejected background galaxies at faint magnitudes bluewards of the main sequence; the prominence of this feature generally correlates with the total area observed, and is accounted for in our CMD modeling procedure as described in the next section.

\begin{figure*}
    \centering
    \includegraphics[width=0.98\textwidth]{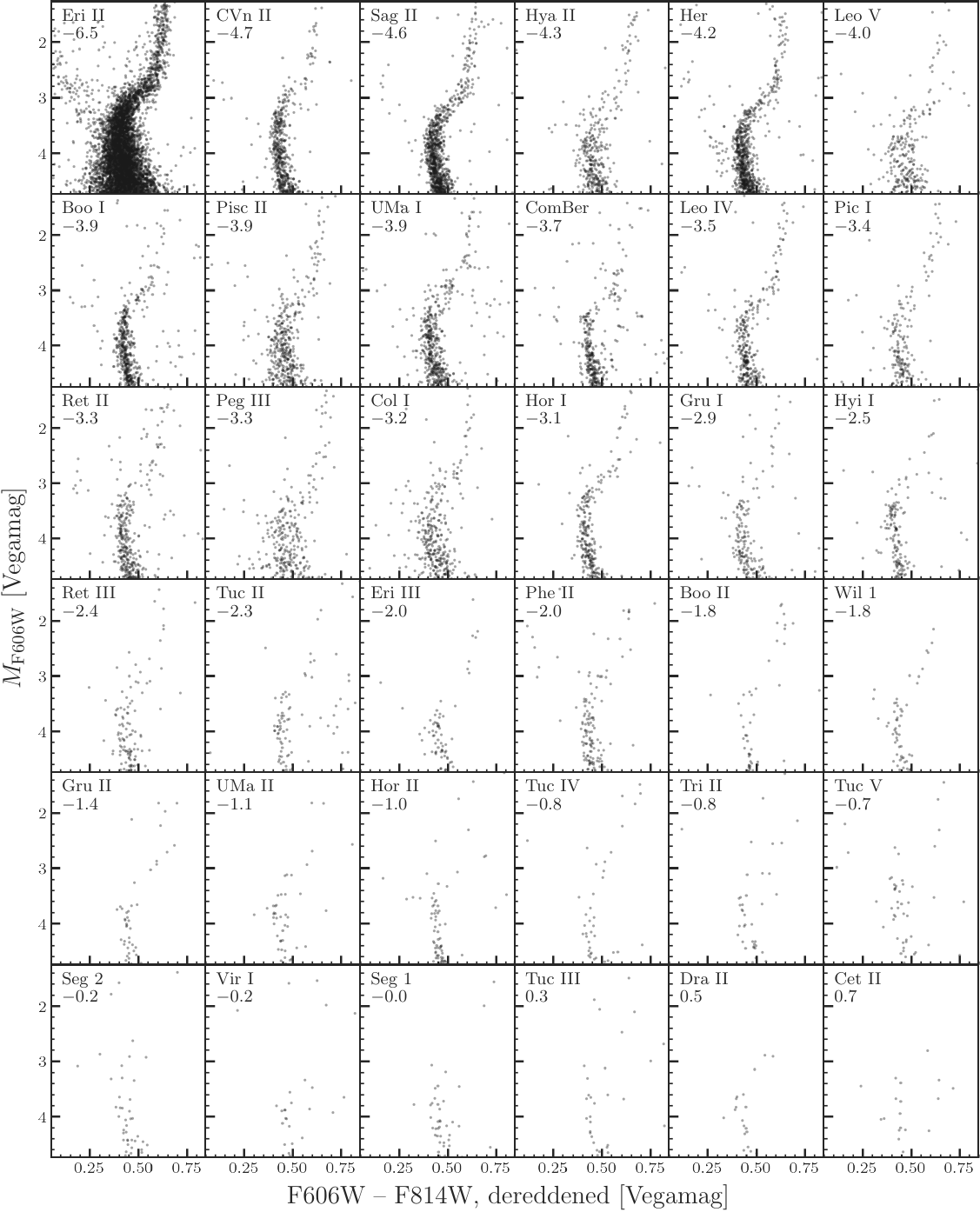}
    \caption{As \autoref{fig:cmds}, but using absolute magnitudes (based on the distances and reddenings reported in \autoref{tab:ssp}) and zooming in on the few magnitudes around the oMSTO, subgiant branch, and base of the red giant branch.}
    \label{fig:abscmds}
\end{figure*}

We reject likely foreground stars based on the membership determinations of \citet{2022A&A...657A..54B}, which combine \emph{Gaia} eDR3 proper motions with spatial and color-magnitude information. We keep only stars with $P_{\mathrm{memb}} \geq 0.75$ for stars with membership probabilities.
We also restrict the input catalogs to stars within two half-light radii of each target using the structural parameters compiled in their Table~B.1.
For Pic~I and Peg~III we extend the selection to 2.5$R_h$, and for Eri~III to 3$R_h$, as we see significant stellar population signal out to these radii in these galaxies.
We additionally reject stars within 2$R_h$ of the globular cluster in Eri~II based on the structural parameters reported by \citet{2021ApJ...908...18S}.

We demonstrate the application of all quality and membership criteria in Hercules in \autoref{fig:cuts}, and show color-magnitude diagrams of the final culled photometry for all galaxies in \autoref{fig:cmds} and \autoref{fig:abscmds}.
\autoref{fig:cmds} shows nearly the full dynamic range of the photometry for all galaxies in apparent magnitudes, and \autoref{fig:abscmds} is restricted to $\pm2$ magnitudes around the MSTO in absolute magnitudes.
We provide the full set of photometric catalogs (e.g., inclusive of all stars regardless of quality, but with columns indicating whether quality criteria are met) as part of our data release.

\subsection{Artificial star tests}

We assess photometric completeness, bias, and scatter for each galaxy using artificial star tests (ASTs), with 100,000 artificial stars per exposure depth per target.
The ASTs are distributed uniformly in color and magnitude, and their locations on the field of view follow the spatial distributions of observed stars. 
Artificial stars are injected into the images one at a time and measured identically to the photometry described in the previous subsection, with the exception of the parameter \texttt{ACSUseCTE}.
Although we used CTE-corrected \texttt{*\_flc} images throughout, for the artificial stars we set \texttt{ACSUseCTE=1} to ensure realistic evaluation of the photometric noise induced by CTE correction, which is not otherwise accounted for in DOLPHOT's AST routine.
All the same photometric quality metrics and spatial culling used on the real catalogs are applied to the artificial star photometry outputs.

\section{Star-formation histories} \label{sec:SFHs}

In this section we present SFHs of all 36 galaxies.
We first measure distance and foreground extinction values for each galaxy via main sequence fitting (detailed in \autoref{ssec:dist_ebv}), and then discuss their SFH fitting process and results (\autoref{ssec:cmd_fits}).
We perform all SFH fitting with \texttt{MATCH} \citep{2002MNRAS.332...91D}, a well-tested software package frequently used in resolved stellar populations studies \citep[e.g.,][]{
2003ApJ...596..253S, 2009AJ....137..419W, 2010ApJ...721..297M, 
2010ApJ...722.1864M, 2011ApJ...739....5W, 
2014ApJ...789..147W, 
2015ApJ...805..183L,
2015ApJ...812..158M, 2017ApJ...837..102S, 
2023ApJ...956...86S, 2025ApJ...979..205S}.
\texttt{MATCH} measures a maximum-likelihood SFH solution by comparing observed and modeled Hess diagrams via Poisson statistics.

\subsection{Distances and foreground extinctions}\label{ssec:dist_ebv}

\begin{figure}
    \centering  
    \includegraphics[width=\linewidth]{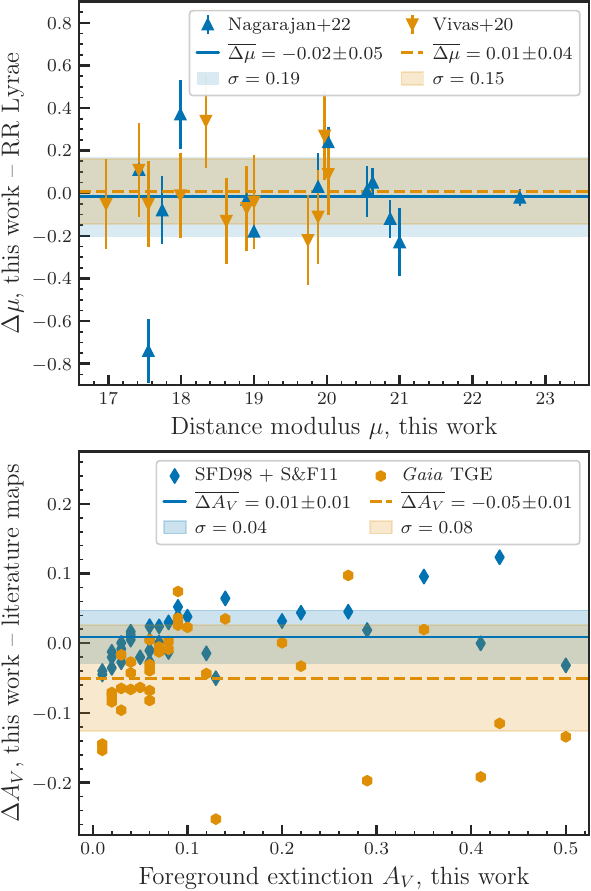}
    \caption{Comparisons of our SSP-based distance and extinction measurements with literature values. Upper panel: comparison of our distance moduli with RR Lyrae results from \citet[][blue triangles]{2022ApJ...932...19N} and \citet[][orange inverted triangles]{2020ApJS..247...35V}. 
    Error bars reflect only the RR Lyrae distances' uncertainties, as we do not calculate formal uncertainties on ours.
    Average differences between respective distance moduli are shown as horizontal lines, and standard deviations as horizontal bars.
    Lower panel: comparison of our $A_V$ with values from the maps of \citet[][blue diamonds]{1998ApJ...500..525S}, with $R_V$ from \citet{2011ApJ...737..103S}, and \citet[][orange hexagons]{2023A&A...674A..31D}.
    As above, averages and standard deviations are marked with horizontal lines and bars.
    }
    \label{fig:distance_comparison}
\end{figure}

\begin{table}
    \centering
    \caption{Best-fit SSP parameters}
    \label{tab:ssp}
\begin{tabular}{lCCCC}
\hline \hline
Galaxy & A_V \rm{\ [mag]} & $\mu$ \rm{\ [mag]} & \log(t/\rm{yr}) & \rm{[Fe/H]} \\
\hline
Boo I & 0.01 & 19.00 & 10.11 & -2.0 \\
Boo II & 0.06 & 18.34 & 10.10 & -2.5 \\
CVn II & 0.04 & 21.00 & 10.13 & -2.8 \\
Cet II & 0.06 & 17.25 & 10.11 & -2.5 \\
Col I & 0.06 & 21.22 & 10.12 & -2.3 \\
ComBer & 0.01 & 17.99 & 10.11 & -2.0 \\
Dra II & 0.09 & 16.58 & 10.12 & -2.1 \\
Eri II & 0.03 & 22.65 & 10.08 & -1.7 \\
Eri III & 0.06 & 19.74 & 10.09 & -2.4 \\
Gru I & 0.04 & 20.55 & 10.12 & -2.6 \\
Gru II & 0.07 & 18.69 & 10.12 & -2.0 \\
Her & 0.22 & 20.63 & 10.13 & -2.6 \\
Hor I & 0.02 & 19.66 & 10.13 & -2.7 \\
Hor II & 0.03 & 19.42 & 10.12 & -2.3 \\
Hya II & 0.20 & 20.68 & 10.12 & -1.6 \\
Hyi I & 0.35 & 17.42 & 10.12 & -2.9 \\
Leo IV & 0.05 & 20.87 & 10.13 & -2.0 \\
Leo V & 0.14 & 21.09 & 10.09 & -1.6 \\
Peg III & 0.41 & 21.25 & 10.09 & -1.5 \\
Phe II & 0.06 & 19.88 & 10.11 & -3.0 \\
Pic I & 0.03 & 20.54 & 10.13 & -2.9 \\
Pisc II & 0.13 & 21.14 & 10.11 & -1.9 \\
Ret II & 0.08 & 17.57 & 10.12 & -3.1 \\
Ret III & 0.12 & 19.97 & 10.10 & -2.5 \\
Sag II & 0.43 & 18.90 & 10.12 & -1.9 \\
Seg 1 & 0.08 & 17.12 & 10.10 & -1.8 \\
Seg 2 & 0.50 & 17.74 & 10.11 & -2.1 \\
Tri II & 0.27 & 17.50 & 10.10 & -2.8 \\
Tuc II & 0.02 & 18.62 & 10.12 & -2.2 \\
Tuc III & 0.02 & 16.97 & 10.09 & -2.7 \\
Tuc IV & 0.04 & 18.41 & 10.11 & -2.6 \\
Tuc V & 0.10 & 18.89 & 10.11 & -2.9 \\
UMa I & 0.09 & 20.02 & 10.10 & -2.9 \\
UMa II & 0.29 & 17.55 & 10.11 & -2.1 \\
Vir I & 0.07 & 19.68 & 10.11 & -1.9 \\
Wil 1 & 0.09 & 17.93 & 10.13 & -3.0 \\
\hline
\end{tabular}
\end{table}

While most, if not all, of our sample have at least one distance measurement available in the literature, they vary greatly in method and precision.
Similarly, foreground extinctions reported by different dust maps \citep[e.g.,][]{1998ApJ...500..525S, 2019ApJ...887...93G, 2022A&A...661A.147L, 2023A&A...674A..31D} differ by up to 0.3 mag in $A_V$ for the same galaxy. The heterogeneity of these values, and their adoption in the literature, can affect the inferred stellar populations of each system.

Therefore, we choose to measure self-consistent distances and extinctions for each galaxy by fitting them as simple stellar populations (SSPs) between the base of the red giant branch and the main sequence knee (approximately 2 $<$ $M_{\mathrm{F606W}}$ $<$ 6 as shown in \autoref{fig:abscmds}).
For this we use the \texttt{MATCH} SSP fitting utility, which performs a grid search over specified ranges of distance, extinction, age, and metallicity for single-age, single-metallicity populations (SSPs).

We use the BaSTI \citep{2004ApJ...612..168P, 2018ApJ...856..125H} stellar model suite \citep{2021ApJ...908..102P} to generate fiducial SSP isochrones. 
We use BaSTI throughout this work because it is the only model suite available in \texttt{MATCH} that covers the extremely low metallicities and alpha abundance ratios found in UFDs.
We choose grids of $10 < \log_{10}(t/\rm{yr}) < 10.14$ (10 - 13.8 Gyr) with 0.01 dex steps and $-3.2 \le$ [Fe/H] $\le -1.4$ with 0.1 dex steps, with [$\alpha$/Fe] $= 0.4$.
For distances we set a range of $\pm$0.4 mag around the literature distance moduli compiled in \citet{2022A&A...657A..54B} with 0.01 mag steps.
For extinction, we take the $A_V$ values reported by the respective all-sky maps of \citet{1998ApJ...500..525S} \citep[recalibrated by][]{2011ApJ...737..103S} and \citet{2023A&A...674A..31D} as starting points, and set the lower limit to 0.05~mag below the minimum of the two values, and the upper limit to 0.05~mag greater than the maximum, with 0.01~mag steps.
We report all best-fit distance moduli $\mu$ and foreground extinctions $A_V$ in \autoref{tab:ssp}, as well as ages and metallicities.

We compare our derived distances to those measured from RR Lyrae variable stars calibrated to \emph{Gaia} 
in the upper panel of \autoref{fig:distance_comparison}.
\citet{2022ApJ...932...19N} recalibrated literature RR Lyrae measurements in 39 Local Group dwarfs, including 13 of the UFDs analyzed here, to anchor them to the \emph{Gaia} Milky Way RR Lyrae period-luminosity-metallicity relation.
\citet{2020ApJS..247...35V} used \emph{Gaia} DR2 data to identify and measure distances to RR Lyrae in 14 nearby UFDs, 12 of which are in our sample.
Of these, 7 galaxies are common between the two studies, for a total of 18 individual galaxies or half our total sample.
We find small mean offsets ($|\overline{\Delta\mu}| \le 0.02$~mag) between our distances and the RR Lyrae results (horizontal lines in \autoref{fig:distance_comparison}), albeit with high scatter ($\sim$0.2~mag, shaded horizontal bars) around the mean.
The outlier at $\Delta\mu = -0.75$ is UMa~II, whose RR Lyrae distance is based on a single RRab star, and for which \citet{2022ApJ...932...19N} note comparable discrepancies between their measurement and prior literature  distances.
In general, our distances are in good agreement with the literature, and have the benefit of uniformity.

\subsection{Foreground \& background populations}

\begin{figure}
    \centering
    \includegraphics[width=\linewidth]{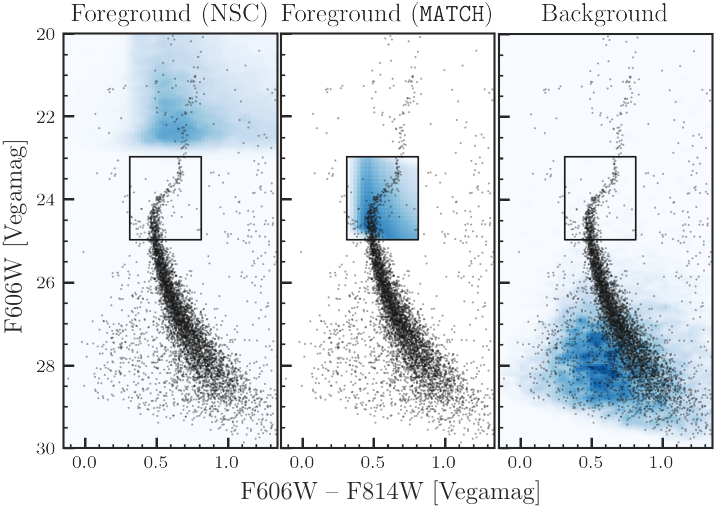}
    \caption{Example foreground and background components for Hercules, shown as blue density maps with scatter plots of the CMD overlaid. The color-magnitude box we use for SFH fitting is outlined in black on each panel. In this case, the NOIRLab Source Catalog (NSC, left panel) is too shallow to reach Hercules' MSTO, so that foreground component has a weight of zero in the resulting SFH fit, and the MATCH foreground model (center panel) is used instead. However, this is not the case for all galaxies. Similarly, the background component (right panel) is negligible here, but in more distant galaxies it carries more weight.}
    \label{fig:fg_bg}
\end{figure}

\texttt{MATCH} offers the option to include components in the likelihood that represent foreground Galactic populations as well as unresolved background galaxies and other non-stellar sources in the CMD fitting process.
In our sparsely populated data, even a small number of spurious sources can affect the CMD fits and resulting SFHs. Therefore, accurately modeling all non-UFD components is essential. 

For the Galactic foreground we supplement the built-in \texttt{MATCH} foreground model, which implements a smooth three-component Milky Way profile \citep{2010ApJ...714..663D}, with $ri$ photometry from Pan-STARRS DR1 \citep[PS1,][]{2016arXiv161205560C, 2020ApJS..251....7F} or the NOIRLab Source Catalog DR2 \citep[NSC,][]{
2018AJ....156..131N, 2021AJ....161..192N} depending on availability.
We query the respective catalogs within an annulus around each galaxy, starting at 6$R_h$ and extending to 10$R_h$ or 0.5$^{\circ}$, whichever is larger.
The PS1 and NSC catalogs reach typical limiting $r$-band magnitudes of approximately 21 and 23 respectively, which cover the oMSTO for only the nearest $\sim$half of the galaxy sample. However, they may be informative as to any coincident Galactic halo substructures \citep[as in the case of, e.g., Gru~II and the Chenab/Orphan Stream,][]{2019MNRAS.490.2183M}.
We convert the $ri$ magnitudes from their native AB system to the Vega system, but do not apply any additional color transformations to put them on the HST filter system, as the background structures are intrinsically diffuse in color-magnitude space over the narrow color range of interest, and convolved with a smoothing kernel in \texttt{MATCH} (see \autoref{ssec:cmd_fits} and \autoref{fig:fg_bg}).

For the background, we use the subset of culled photometry that meets the per-filter GST quality cuts but fails the global quality cuts. These largely consist of unresolved background galaxies and shredded diffraction spikes from saturated foreground stars.
We allow a linear scaling factor for all foreground and background components to vary freely in the CMD fitting.
We illustrate all foreground and background components against the final fitted CMD for Hercules in \autoref{fig:fg_bg}.

\subsection{CMD fits} \label{ssec:cmd_fits}

\begin{figure}
    \centering
    \includegraphics[width=\linewidth,trim=0 0 0 0.5cm,clip=True]{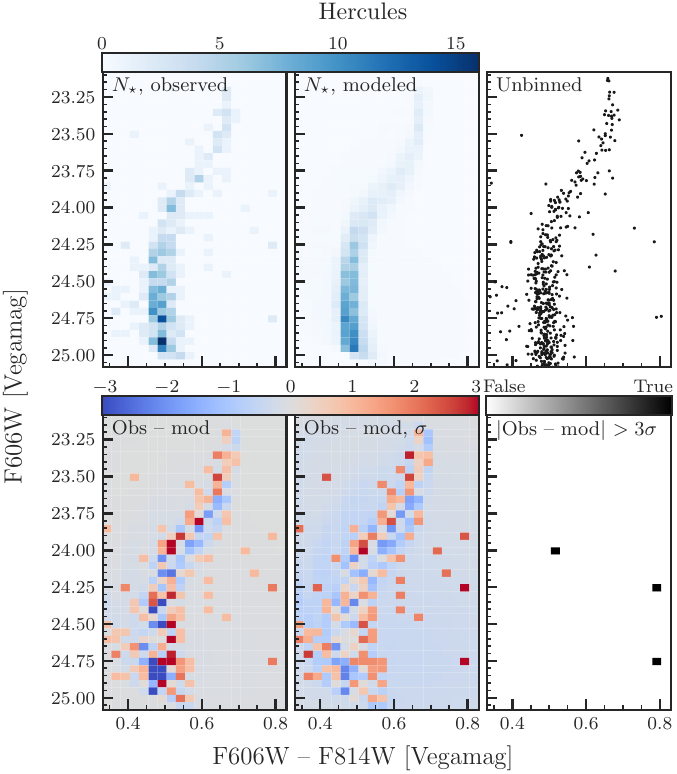}
    \caption{Example MSTO fitting for Hercules. Upper row, left to right: observed and modeled Hess diagrams, and the unbinned CMD for reference. Lower row: Hess diagram residuals (observed -- modeled) and residual significance, and bins with residual significance over 3$\sigma$.
    The complete figure set (36 images) is available in the online journal.
    }
    \label{fig:match_cmds}
\end{figure}

With distances, extinctions, and all foreground and background components in hand, we generate stellar population models with ages between $9.78 \le \log_{10}(t/\rm{yr}) < 10.14$ (6 -- 13.8 Gyr) with an 0.01 dex step size, and metallicities between $-3.2 <$ [Fe/H] $< -1.2$ with an 0.1 dex step, again using the BaSTI model suite with $[\alpha/\rm{Fe}]=0.4$.
We use a Kroupa initial mass function \citep{2001MNRAS.322..231K} normalized between 0.08 $M_{\odot}$ and 120 $M_{\odot}$, 
an unresolved binary fraction of 0.35, and bin sizes of 0.025, 0.05 mag in color and magnitude respectively. 
Following previous analyses of Local Group dwarfs \citep[e.g.,][]{2014ApJ...789..148W, 2017ApJ...837..102S, 2023ApJ...956...86S}, we impose an age-metallicity relation constraint using the \texttt{-zinc} option, which specifies that the mean metallicity must increase monotonically over time.
This option mitigates the age-metallicity degeneracy at the oMSTO, particularly for the metallicity-insensitive F606W-F814W filter combination \citep{2023ApJ...956...86S}. 
We allow the initial metallicity to vary between --3.2~$\leq$~[Fe/H]~$\leq$~--2.0 dex, and the final between --3.2~$\leq$~[Fe/H]~$\leq$~--1.6 dex.
We additionally set a metallicity dispersion of $\langle\sigma_{\rm{[Fe/H]}}\rangle=0.2$~dex per age bin \citep{2011ApJ...727...78K,2018MNRAS.474.2194E}.

We set the default magnitude range for the CMD fitting to within $\pm$1~mag of the oMSTO ($2.5 \leq M_{\rm{F606W}} \leq 4.5$); %
for galaxies with fewer than 100 stars within this range, we extend it to $\pm$2~mag around the oMSTO.
For Eri~II, which is both the most populated and most distant of all galaxies in our sample, we restrict the fits to $\pm$0.75~mag around the oMSTO; this helps reduce errors from low-S/N sources at the faint end.
We show an example set of Hess diagrams from the CMD modeling process for Hercules in \autoref{fig:match_cmds}.  The model provides an excellent match to the data, with few points discrepant at a level $>3\sigma$. 

\begin{figure*}
    \centering
    \includegraphics[width=\linewidth]{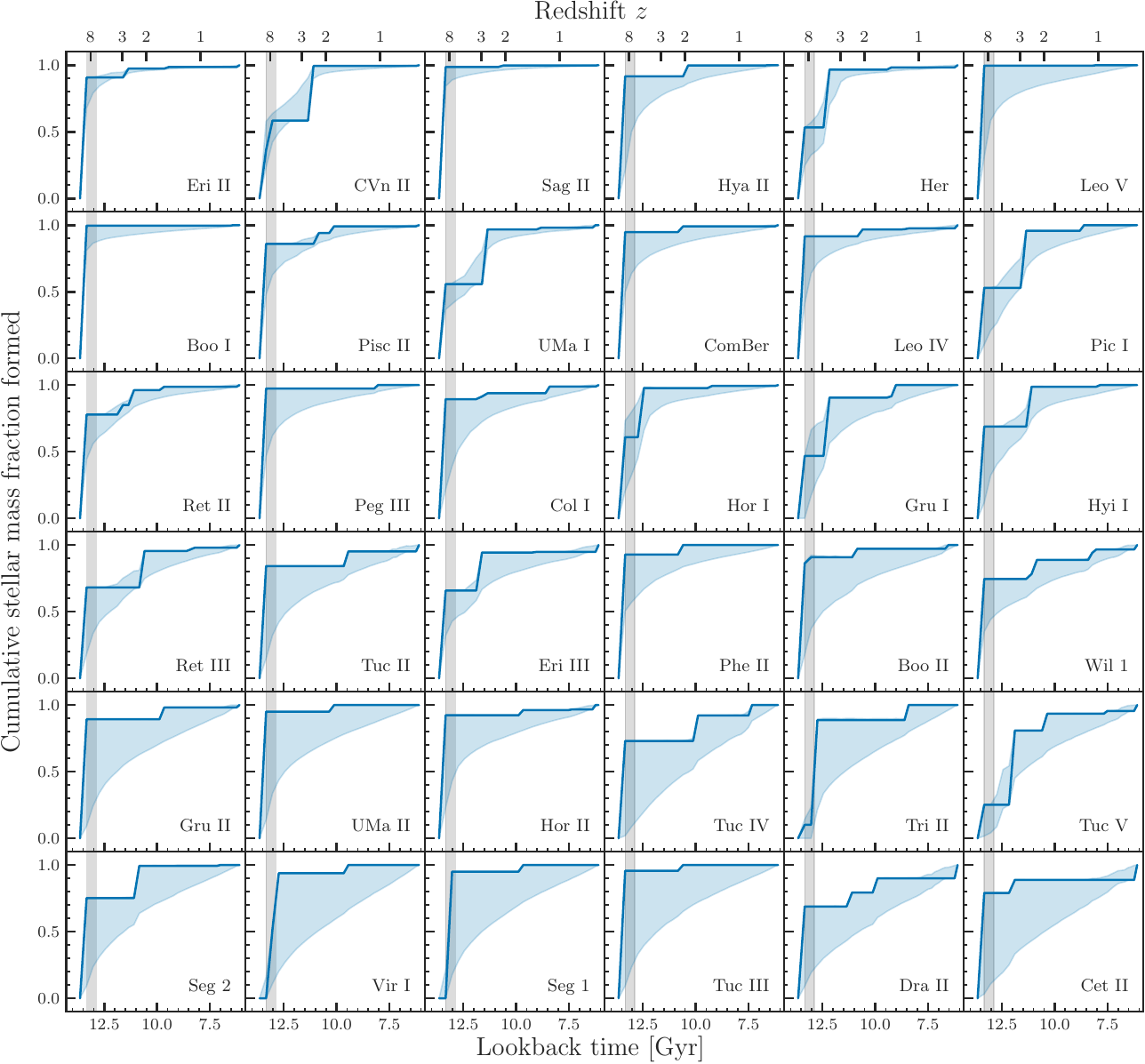}
    \caption{Best-fit SFHs for all UFDs (blue lines) with 68\% confidence intervals (light blue shaded regions). The epoch of reionization is marked with the gray vertical bands \citep[12.87 -- 13.33 Gyr ago in $\Lambda$CDM;][]{2020A&A...641A...6P, 2021MNRAS.505.2764B}. 
    Uncertainties are from random errors only, and do not include systematic uncertainties on models or on galaxy parameters such as distance and reddening.}
    \label{fig:sfhs}
\end{figure*}

Once the best-fit SFHs are in hand, we calculate the statistical uncertainties on them using a Hamiltonian Monte Carlo algorithm \citep{1987PhLB..195..216D}, as described in \citet{2013ApJ...775...76D}.
We do not compute systematic uncertainties, which are designed to capture shortcomings of stellar evolution models \citep{2012ApJ...751...60D}, as we are primarily interested in differences between the SFHs in a relative sense.
We show all cumulative SFH fits with 68\% confidence intervals in \autoref{fig:sfhs}.

\subsection{Quenching timescales}

The epoch at which a galaxy ceases forming stars, or quenches, is an important quantity for understanding the physical processes that drive its lifetime evolution, such as feedback, environmental effects, and---in the case of UFDs---reionization.
However, its interpretation from CMD-based SFH measurements is rarely straightforward.
Often, a metric is adopted that describes the lookback time by which some fraction of the total stellar mass has formed, denoted $\tau_{q}$, where $q$ is a percentage \citep[e.g.,][]{2014ApJ...789..148W, 2015ApJ...804..136W, 2017ApJ...837..102S}.

While $\tau_{90}$ is the typical choice for more massive galaxies, UFDs are especially vulnerable to inflated estimates of late-time star formation from small numbers of spurious sources, such as unrejected foreground main sequence stars or blue stragglers,
both of which can mimic younger stellar populations due to their CMD locations.
We attribute the low levels of relatively recent star formation ($\lesssim 10$~Gyr ago) found in Cet~II, Dra~II and Wil~I to such spurious populations (see \autoref{fig:abscmds}, \autoref{fig:sfhs}, and \autoref{tab:quench}). \added{Each of these have several stars near the oMSTO that are slightly too blue to belong to the dominant population, such that they are interpreted by the SFH fitting software as younger populations; however, it is unclear if these stars are truly young, or if they are blue stragglers or non-members.}

Therefore, we prefer the more conservative $\tau_{80}$ as our fiducial quenching time, following \citet{2023ApJ...956...86S} and \citet{2024ApJ...967..161M}.
Using this definition, we find an average quenching time of $12.48 \pm 0.18$ Gyr ago ($z = 4.6_{-0.5}^{+0.6}$ in $\Lambda$CDM), broadly consistent with previous studies \citep{2014ApJ...796...91B, 2021ApJ...920L..19S} and with the end of the epoch of reionization in $\Lambda$CDM \citep{2020A&A...641A...6P, 2021MNRAS.505.2764B}.
(Note that this value includes Sag~II; excluding it shifts the average lookback time to 12.45~Gyr.)
We report all $\tau_{80}$ values in \autoref{tab:quench}, and include $\tau_{50}$ and $\tau_{90}$ as well for completeness. 
In the remainder of the analysis, we present results for both $\tau_{80}$ and $\tau_{90}$, but take $\tau_{80}$ as the fiducial quenching epoch.
All reported $\tau$ values and their uncertainties are derived by interpolating the cumulative star formation histories shown in \autoref{fig:sfhs}; $\tau_{80}$, for example, is the best-fit value (dark blue lines) at a cumulative stellar mass fraction of 0.8, and the upper and lower uncertainties are the limits of the 68\% confidence window (light blue patches) at the same mass fraction.

\begin{table*}
    \centering
    \caption{Stellar mass formation lookback times}
    \label{tab:quench}
\begin{tabular}{llRRR}
\hline \hline
Galaxy & Kinematic group & \tau_{50}\ \rm{[Gyr]} & \tau_{80}\ \rm{[Gyr]} & \tau_{90}\ \rm{[Gyr]} \\
\hline
Boo I & Long-term MW & 13.49_{-0.04}^{+0.00} & 13.40_{-0.06}^{+0.00} & 13.37_{-1.03}^{+0.00} \\
Boo II & Long-term MW & 13.47_{-0.99}^{+0.00} & 13.36_{-3.92}^{+0.00} & 13.09_{-5.56}^{+0.08} \\
CVn II & First infall & 13.15_{-0.44}^{+0.23} & 11.21_{-0.04}^{+0.63} & 11.15_{-0.07}^{+0.23} \\
Cet II & Long-term MW & 13.45_{-2.96}^{+0.00} & 12.13_{-4.66}^{+0.02} & 6.22_{-0.00}^{+1.77} \\
Col I & Long-term MW & 13.47_{-0.70}^{+0.00} & 13.37_{-2.64}^{+0.00} & 11.82_{-3.33}^{+0.00} \\
ComBer & Long-term MW & 13.48_{-0.15}^{+0.00} & 13.38_{-1.57}^{+0.00} & 13.35_{-3.69}^{+0.00} \\
Dra II & Long-term MW & 13.42_{-2.65}^{+0.00} & 10.10_{-2.43}^{+0.00} & 9.89_{-3.30}^{+0.00} \\
Eri II & First infall & 13.48_{-0.06}^{+0.00} & 13.37_{-0.40}^{+0.00} & 13.34_{-1.23}^{+0.00} \\
Eri III & Long-term MW & 13.41_{-1.01}^{+0.00} & 11.75_{-1.78}^{+0.05} & 11.66_{-4.19}^{+0.02} \\
Gru I & Long-term MW & 12.43_{-0.16}^{+0.86} & 12.23_{-2.34}^{+0.11} & 12.17_{-3.75}^{+0.01} \\
Gru II & Long-term MW & 13.47_{-1.58}^{+0.00} & 13.37_{-4.66}^{+0.00} & 9.87_{-2.55}^{+0.00} \\
Her & Long-term MW & 13.36_{-0.99}^{+0.00} & 12.27_{-0.46}^{+0.09} & 12.21_{-0.80}^{+0.04} \\
Hor I & LMC & 13.39_{-0.73}^{+0.04} & 12.59_{-0.63}^{+0.45} & 12.51_{-2.27}^{+0.16} \\
Hor II & Long-term MW & 13.48_{-0.63}^{+0.00} & 13.38_{-3.53}^{+0.00} & 13.34_{-5.56}^{+0.00} \\
Hya II & First infall & 13.48_{-0.44}^{+0.00} & 13.38_{-2.08}^{+0.00} & 13.34_{-3.67}^{+0.00} \\
Hyi I & LMC & 13.42_{-1.16}^{+0.00} & 11.25_{-0.50}^{+0.26} & 11.17_{-2.14}^{+0.07} \\
Leo IV & First infall & 13.48_{-0.26}^{+0.00} & 13.38_{-1.44}^{+0.00} & 13.34_{-3.50}^{+0.00} \\
Leo V & Long-term MW & 13.49_{-0.36}^{+0.00} & 13.40_{-1.82}^{+0.00} & 13.37_{-4.13}^{+0.00} \\
Peg III & Long-term MW & 13.49_{-0.22}^{+0.00} & 13.39_{-1.63}^{+0.00} & 13.36_{-3.76}^{+0.00} \\
Phe II & LMC & 13.48_{-0.14}^{+0.00} & 13.38_{-2.30}^{+0.00} & 13.35_{-4.32}^{+0.00} \\
Pic I & Long-term MW & 13.35_{-1.57}^{+0.00} & 11.45_{-0.96}^{+0.11} & 11.39_{-2.65}^{+0.04} \\
Pisc II & Long-term MW & 13.47_{-0.17}^{+0.00} & 13.36_{-1.72}^{+0.00} & 10.97_{-0.78}^{+0.39} \\
Ret II & LMC & 13.45_{-0.26}^{+0.00} & 11.80_{-0.48}^{+0.50} & 11.24_{-1.29}^{+0.07} \\
Ret III & Long-term MW & 13.42_{-1.19}^{+0.00} & 10.73_{-0.74}^{+0.41} & 10.64_{-2.44}^{+0.04} \\
Sag II & Long-term MW & 13.49_{-0.03}^{+0.00} & 13.39_{-0.04}^{+0.00} & 13.36_{-0.52}^{+0.00} \\
Seg 1 & Long-term MW & 13.18_{-1.56}^{+0.04} & 13.08_{-4.36}^{+0.01} & 13.05_{-5.69}^{+0.00} \\
Seg 2 & Long-term MW & 13.44_{-1.87}^{+0.00} & 11.04_{-2.16}^{+0.01} & 10.94_{-3.48}^{+0.01} \\
Tri II & Long-term MW & 12.88_{-1.44}^{+0.03} & 12.77_{-4.36}^{+0.01} & 8.59_{-1.35}^{+0.02} \\
Tuc II & Long-term MW & 13.46_{-1.12}^{+0.00} & 13.35_{-4.24}^{+0.00} & 9.55_{-2.16}^{+0.01} \\
Tuc III & Long-term MW & 13.48_{-1.67}^{+0.00} & 13.39_{-4.52}^{+0.00} & 13.35_{-5.90}^{+0.00} \\
Tuc IV & Long-term MW & 13.43_{-3.12}^{+0.00} & 10.03_{-2.34}^{+0.02} & 9.91_{-2.78}^{+0.00} \\
Tuc V & Long-term MW & 12.04_{-0.48}^{+0.43} & 11.89_{-3.42}^{+0.00} & 10.42_{-3.43}^{+0.00} \\
UMa I & First infall & 13.37_{-1.05}^{+0.00} & 11.46_{-0.09}^{+0.19} & 11.39_{-0.96}^{+0.07} \\
UMa II & Long-term MW & 13.48_{-1.03}^{+0.00} & 13.39_{-4.04}^{+0.00} & 13.35_{-5.66}^{+0.00} \\
Vir I & Long-term MW & 13.05_{-2.13}^{+0.01} & 12.83_{-4.74}^{+0.00} & 12.76_{-5.71}^{+0.00} \\
Wil 1 & Long-term MW & 13.44_{-0.81}^{+0.00} & 11.05_{-2.26}^{+0.00} & 8.37_{-0.93}^{+0.08} \\
\hline
\end{tabular}
\end{table*}

\section{Discussion} \label{sec:discussion}

\subsection{Recommendations for robust MSTO fitting}

\begin{figure*}
    \centering
    \includegraphics[width=0.8\linewidth]{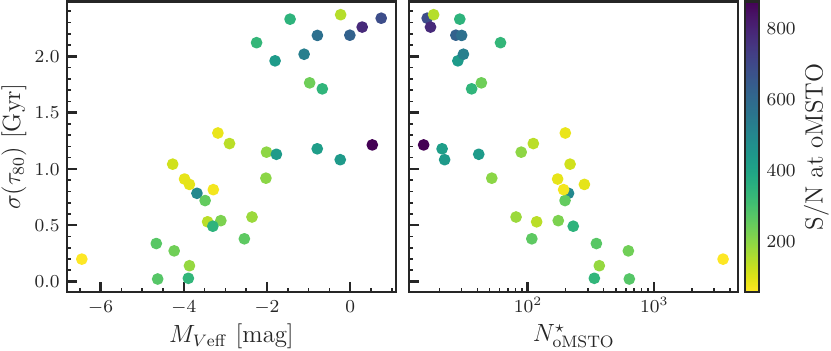}
    \caption{Error on $\tau_{80}$ (half the width of the 68\% confidence interval) against the effective $V$ magnitude of the \HST observations, $M_{V\rm{eff}} = M_V - 2.5\log_{10}(f_\star)$ (left), and the number of stars within $\pm1$~mag of the F606W oMSTO (right;  $2.5 < M_{\rm{F606W}}<4.5$), colored by S/N at the F606W oMSTO. From these we infer $M_{V\rm{eff}} \leq -2.5$ or 100 oMSTO stars as the limits for which one can derive precise SFHs, with S/N as a minor secondary factor.}
    \label{fig:mveff_t80_snr}
\end{figure*}

Our SFHs vary greatly in precision, largely as a function of intrinsic luminosity and of total observed area relative to the galaxy profile.
In \autoref{fig:mveff_t80_snr} we examine the $1\sigma$ width of the uncertainty on $\tau_{80}$ as a function of the effective $V$ magnitude $M_{V\rm{eff}}$ of the \HST observations as defined in \autoref{sec:obs}, and of the number of stars within $\pm1$ magnitude of the oMSTO.
The marker colors represent the signal-to-noise at the oMSTO ($M_{\rm{F606W}}=3.5$).

We find that a well-sampled stellar population---both in terms of the observational footprint and the galaxy's total stellar mass itself---vastly outweighs photometric S/N for the purposes of achieving precise CMD-based SFHs, at least in the high-S/N limit of our data.
Although Eri~II has the lowest S/N at the oMSTO ($\sim$50), it is also by far the best populated;
conversely, Dra~II has the highest S/N---nearly 900---near the oMSTO, but simply has too few stars to meaningfully discriminate among isochrones.

Crucially, these results demonstrate that there is a lower bound on not only the effective luminosity for which oMSTO fitting can reasonably be applied, but on the intrinsic luminosity of the galaxy as well, indicating a stellar mass threshold below which CMD fitting ceases to be effective as a means of SFH determination.
Of course, this threshold is dependent on the acceptable uncertainties for a given science case. 
For the purposes of testing patchy reionization, we prefer uncertainties on $\tau_{80}$ below 500~Myr, so that an average of at least three galaxies can achieve an uncertainty less than half of the predicted $\sim$500-600~Myr time lag in quenching \citep{2012ApJ...746..109L, 2014ApJ...785..134L,2018ApJ...856L..22A,2019ApJ...882..152Z, 2020MNRAS.494.2200K, 2020MNRAS.496.4087O, 2023ApJ...959...31K}. 
The faintest galaxy at which we reach this limit is Hyi~I at $M_{V\rm{eff}} = -2.5$; this corresponds to $\sim100$ stars within $\pm1$~mag of the oMSTO, consistent with the recommendation of \citet{2014ApJ...796...91B}.
More lax uncertainty thresholds (e.g., 1~Gyr) may be acceptable for the purpose of measuring ensemble quenching timescales.

For the galaxies in our sample, observations over a larger area (e.g., with the upcoming Nancy Grace Roman Space Telescope) would be most beneficial in the cases of UMa~II, Tuc~IV, and Gru~II, all of which have $M_V \le -3.5$ but $f_{\star} \le 0.1$ ($M_{V\rm{eff}} \ge -1.9$).
Improving the S/N at the oMSTO, despite its generally marginal effects at both the well- and sparsely-populated extremes, would be most useful in Col~I, Leo~V, Peg~III, and Pisc~II, which have S/N$_{\rm MSTO}$ $<$ 100.
These are otherwise well-sampled populations for which star-galaxy separation near the turnoff is poorest, which may have the effect of biasing the SFHs or inflating their uncertainties (see \autoref{fig:cmds} and \autoref{fig:sfhs}).

Additional potential avenues for improvement in future studies include: a) robust membership information at fainter magnitudes, via deeper proper motion data (e.g., with future \gaia data releases or Vera C. Rubin observations) and/or velocities (e.g.\ with ELTs); and b) use of a filter pair with a broader color baseline.
We stress that the F606W--F814W filter combination is suboptimal for SFH fitting due to its narrow color separation, where sensitivity to metallicity variations is minimal.
The F475W--F814W combination on ACS/WFC and WFC3/UVIS offers much greater leverage on color and thus metallicity \citep[as demonstrated by][]{2025ApJ...979..205S}, but at greater observational cost. The F475W--F814W  combination is only available in the \HST\ archive for 17 of the galaxies in this study as of this writing.
This is less than half of our sample, and the majority of the F475W observations are much shallower than the F606W, targeting the red giant branch rather than the MSTO.

\subsection{Comparison with literature}

\begin{figure*}
    \centering
    \includegraphics[width=0.8\linewidth]{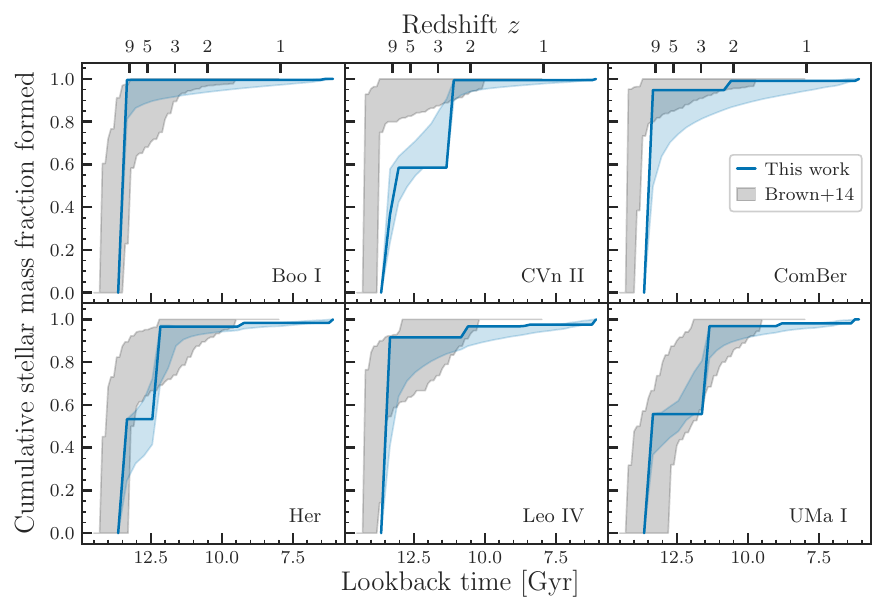}
    \caption{Comparison of the SFHs derived in this work (blue) with the SFHs reported by \citet[][gray patches]{2014ApJ...796...91B}.
    We find excellent agreement in most cases, with only minor exceptions.}
    \label{fig:brown14_comparison}
\end{figure*}

\begin{figure*}
    \centering
    \includegraphics[width=\linewidth]{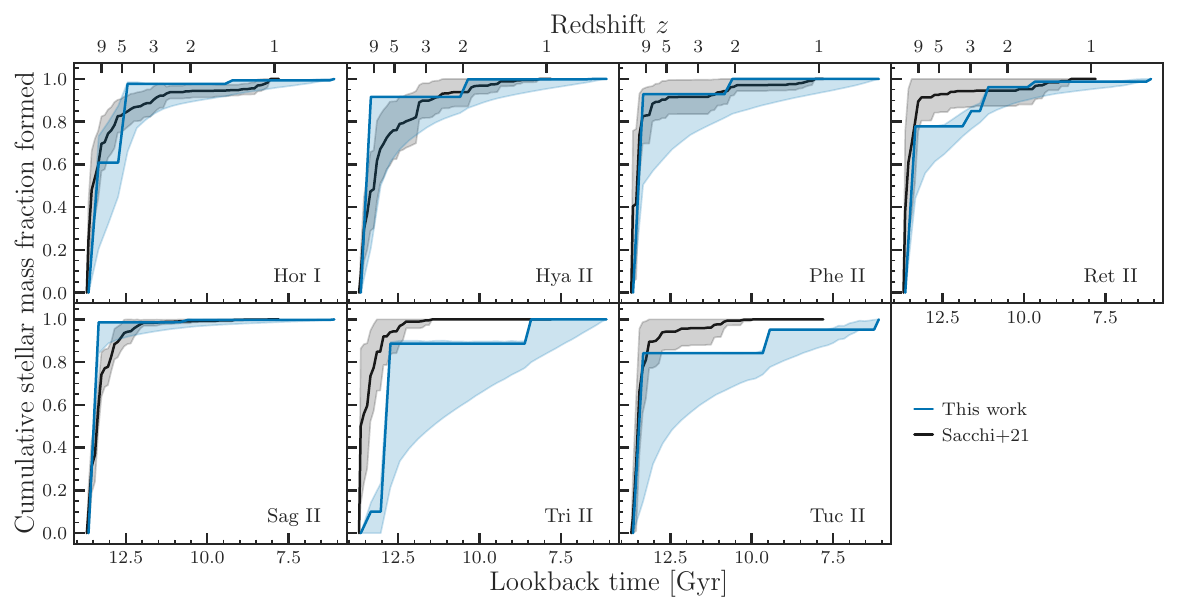}
    \caption{As \autoref{fig:brown14_comparison}, with the SFHs reported by \citet{2021ApJ...920L..19S} as black lines.
    Again, we see generally excellent agreement with some exceptions. The two galaxies with the strongest disagreement, Tri II and Tuc II, have large error bars in our fits.
    }
    \label{fig:sacchi21_comparison}
\end{figure*}

In this section, we undertake a comparison between the 14 MW UFDs in the literature that have SFHs and those presented in this paper. Two papers, \citet{2014ApJ...796...91B} and \citet{2021ApJ...920L..19S}, contain virtually all literature UFD SFHs.  Accordingly, we focus most of our comparisons on these two systematic studies. 

\citet{2014ApJ...796...91B} reported SFHs of six UFDs (Boo~I, CVn~II, Com~Ber, Her, Leo~IV, and UMa~I). They used custom Victoria-Regina isochrones \citep{2014ascl.soft04010V, 2014ApJ...794...72V} along with spectroscopic metallicity distribution functions of red giants measured from Keck spectra, and theoretical [O/Fe] abundances varying as a function of [Fe/H].
SFHs were measured as two-burst models, with absolute ages set relative to a fiducial SSP model of M92.
Distances and foreground reddenings to each galaxy were included as free parameters in the fits.

\citet{2021ApJ...920L..19S} presented SFHs of seven UFDs: Hor~I, Hya~II, Phe~II, Ret~II, Sag~II, Tri~II, and Tuc~II.
They use the SFERA software \citep{2015ApJ...811...76C} with two sets of isochrones: Victoria-Regina with [$\alpha$/Fe] $= +0.4$ \citep{2014ascl.soft04010V, 2014ApJ...794...72V}, and scaled-solar MIST \citep{2016ApJS..222....8D}.

As shown in \autoref{fig:brown14_comparison}, our SFHs are generally in good agreement with those reported by \citet{2014ApJ...796...91B}, modulo respective choices of oldest age limit (14.1 vs.\ 13.7 Gyr). Nearly all galaxies exhibit similar qualitative SFH shapes and overlapping uncertainties. 
Only CVn~II shows any noticeable discrepancy above the $1\sigma$ level at lookback times greater than $\sim12$~Gyr.  
Further, our SFH for CVn~II has a quenching time $>10$~Gyr ago, like \citet{2014ApJ...796...91B}.  
This is older than the $\sim8$~Gyr quenching epoch reported by \citet{2014ApJ...789..147W}, which used lower S/N HST/WFPC2 data and a different distance/extinction combination to measure the SFH.

We note that our distance modulus to Hercules is $\sim0.3$~mag closer than that of \citet{2014ApJ...796...91B} (20.63 vs.\ 20.92), and Leo IV is 0.25~mag closer (20.87 vs.\ 21.12), but the rest are within $\pm 0.1$~mag of each other.
The extinctions show small offsets, with ours systematically lower than those of \citet{2014ApJ...796...91B} by $\sim0.05$-0.1~mag.

The comparison with \citet{2021ApJ...920L..19S} is for the most part favorable, with minor exceptions.
As illustrated in \autoref{fig:sacchi21_comparison}, Hor~I, Hya~II, Phe~II, and Sag~II all have overlapping uncertainties at more or less all ages.
For Ret~II we find evidence of a slightly younger secondary population ($\sim$20\% by mass, formed between 12.5-11~Gyr ago) that is absent in \citet{2021ApJ...920L..19S}.
We find Tri~II and Tuc~II to be somewhat younger than \citet{2021ApJ...920L..19S}, but the large error bars on our results for these two galaxies indicate that these differences should not be over-interpreted.

There is also generally good agreement between our SFHs and the handful of single galaxy SFH papers in the literature.  For example, \citet{2023ApJ...944...43S} measure the SFH of Ret~II using the same 12 HST fields as our analysis (compared to the single field of Ret~II analyzed by \citealt{2021ApJ...920L..19S}).  Our SFH of Ret~II is consistent with that of \citet{2023ApJ...944...43S}, who also find that up to 20\% of the stellar mass formed in a secondary burst $\sim11$~Gyr ago.  

There is modest disagreement in the literature over the SFH of Eri~II.  Our SFH shows that Eri II formed 90\% of its stellar mass prior to $\sim13$~Gyr ago and the remaining 10\% between 12.5 and $\sim9$~Gyr ago.  \citet{2021ApJ...908...18S} analyze the same deep HST data and find that 80\% of the star formation occurred by $\sim13$~Gyr ago, with the remainder concluding by no later than $\sim10$~Gyr ago.  Using multiple deep HST datasets, \citet{2021ApJ...909..192G} find that star formation in Eri~II likely entirely concluded by $\sim13$~Gyr, and that the low-level star formation, $\sim20$\% of the total stellar mass, that extends to $\sim9$~Gyr ago is consistent with zero star formation at these younger ages.  Finally, \citet{2023ApJ...948...50W} use archival HST imaging to measure the SFH of Eri~II, and its globular cluster, and find similar solutions: $\sim70$\% of its stellar mass formed by $\sim13$~Gyr ago and 20\% formed 11-12~Gyr ago.  The main differences among these solutions appear to the significance attributed to any star formation younger than $\sim13$~Gyr.  \citet{2021ApJ...909..192G} and \citet{2021ApJ...908...18S} both conclude that the younger episodes of star formation are of low significance, whereas \citet{2023ApJ...948...50W} and the present paper suggest the younger star formation is significant at least at the 68\% confidence level.  The differences in analysis techniques, adopted oldest ages, and varying uncertainty computations makes it challenging to directly compare these results.  However, we emphasize that the studies are generally telling the same story: Eri~II is a pre-reionization fossil with marginal evidence of extended star formation to no more than $\sim11$~Gyr ago.

As part of developing a new SFH fitting code, \citet{2024arXiv240719534G} measure the SFH of Hor~I using the same HST imaging as in this paper.  They report a SFH that is marginally younger than that of \citet{2021ApJ...920L..19S}, which is similar to our SFH of this galaxy.

\subsection{Environmental quenching differences}

\begin{figure*}
    \centering
    \includegraphics[width=0.9\linewidth]{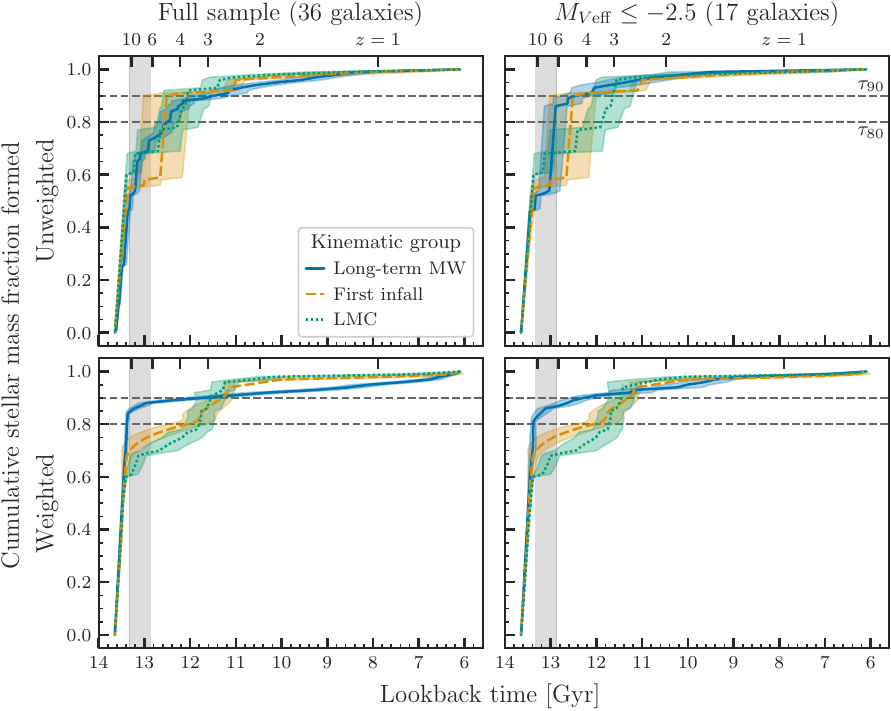}
    \caption{Average SFHs by kinematic group for our full galaxy sample (left column) and galaxies with $M_{V\rm{eff}}$ less than $-2.5$ (right column), computed as simple averages (upper row) and inverse variance-weighted averages (lower row). 
    We see significant differences in $\tau_{80}$ and $\tau_{90}$ (black dashed lines) between the long-term MW sample and the other two kinematic groups only in the weighted averages, whereas the unweighted averages show largely marginal differences.
    }
    \label{fig:kinematic_group_sfhs}
\end{figure*}

In patchy reionization scenarios, the progenitors of MW- and M31-sized galaxies, which were dispersed throughout the volume of the proto-Local Group, as well as progenitors of their present day satellites, may have experienced reionization at different times.  The timing and amplitude of reionization are thought to be functions of local galaxy density, as well as internal factors such as escape fraction of ionizing photons.  Because of their low halo masses, star formation in UFDs at the time of reionization is thought to be particularly sensitive to the impacts of reionization.  Indeed, several simulations suggest that patchy reionization can lead to spreads in the times at which reionization impacts (i.e., quenches) the lowest-mass galaxies by up to 400-600~Myr depending on their environment at the time of reionization \citep[e.g.,][]{2013MNRAS.432.1989S,2018ApJ...856L..22A,2023ApJ...959...31K}.  
\added{These time differences are typically measured across many progenitor halos over volumes thought to be comparable to the ancient LG environment. However, the surviving satellites of a single MW-mass host halo may still bear imprints of patchy reionization, depending on their proximity to the nearest massive halo during reionization and the spatial scales over which massive halos drive early reionization \citep[e.g.,][]{2015ApJ...807...49W,2025arXiv250716245Z}. 
}

We consider whether such signatures are present in our data by comparing the SFHs of UFDs that have different kinematic memberships (e.g., MW satellites vs. LMC satellites vs. first-infall) as determined by the best-availble orbital histories to date.  These different groups may proxy different environmental conditions in the early Universe. Though the exact locations of UFDs at the time of reionization are unknown, the orbital histories provide the best available empirical constraints on their likely locations (e.g., higher vs lower density regions) in the early Universe.  We discuss caveats implicit in this assumption below.

\begin{table*}
    \centering
    \caption{Average $\tau_{80}$ and $\tau_{90}$ values and differences by kinematic group}
\begin{tabular}{cccccc}
\hline \hline
Group & $N$ & $\tau_{80}$ [Gyr] & $\tau_{90}$ [Gyr] & $\Delta \tau_{80}$, MW -- X & $\Delta \tau_{90}$, MW -- X \\ \hline
\multicolumn{6}{c}{Unweighted} \\
Long-term MW & 9 (27) & 12.92$\pm$0.25 (12.49$\pm$0.21) & 12.44$\pm$0.32 (11.49$\pm$0.37) & --- & --- \\
First infall & 5 (5) & 12.56$\pm$0.50 (12.56$\pm$0.56) & 12.51$\pm$0.51 (12.51$\pm$0.51) & 0.36$\pm$0.56 (--0.07$\pm$0.54) & --0.07$\pm$0.60 (--1.02$\pm$0.63) \\
LMC & 3 (4) & 11.88$\pm$0.39 (12.26$\pm$0.46) & 11.64$\pm$0.44 (12.06$\pm$0.53) & 1.04$\pm$0.46 (0.23$\pm$0.51) & 0.80$\pm$0.54 (--0.57$\pm$0.65) \\
\multicolumn{6}{c}{Weighted} \\
Long-term MW & 9 (27) & 13.38$\pm$0.05 (13.38$\pm$0.03) & 12.28$\pm$0.29 (11.73$\pm$0.39) & --- & --- \\
First infall & 5 (5) & 12.05$\pm$0.41 (12.05$\pm$0.41) & 11.30$\pm$0.23 (11.30$\pm$0.23) & 1.33$\pm$0.41 (1.33$\pm$0.41) & 0.98$\pm$0.37 (0.43$\pm$0.45) \\
LMC & 3 (4) & 11.73$\pm$0.31 (11.81$\pm$0.32) & 11.45$\pm$0.29 (11.55$\pm$0.33) & 1.65$\pm$0.31 (1.57$\pm$0.32) & 0.83$\pm$0.41 (0.18$\pm$0.51) \\ 

\hline
\end{tabular}
    \label{tab:t80_by_group}
\tablecomments{We present values for the culled galaxy sample first, and for the full sample in parentheses after. The culled sample is restricted to galaxies with $M_{V\rm{eff}} \le -2.5$, and excludes Sag~II, as in the second column of \autoref{fig:kinematic_group_sfhs}.}
\end{table*}

In \autoref{fig:kinematic_group_sfhs} we show combined SFHs of all 36 galaxies (upper panel) and the 17 galaxies with $M_{V\rm{eff}} \le -2.5$ (see \autoref{fig:mveff_t80_snr}), aggregated by kinematic group.
We exclude Sag~II from the latter due to its status as a likely globular cluster \citep{2021MNRAS.503.2754L}.
We assign galaxies to kinematic groups based on recent work on their orbital histories \citep{2018A&A...619A.103F,2018ApJ...867...19K,2019arXiv190604180F,2020ApJ...893..121P,2022ApJ...940..136P}.
We consider Hor~I, Hyi~I, Phe~II, and Ret~II to be LMC satellites \citep[either long-term or recently captured;][]{2020ApJ...893..121P}, and CVn~II, Eri~II, Hya~II, Leo~IV, and UMa~I to be on their first infall into the MW halo.
We consider the rest of the sample to be long-term MW satellites by default. Of the galaxies with $M_{V\rm{eff}} \le -2.5$, this includes Boo~I, Col~I, ComBer, Gru~I, Her, Leo~V, Peg~III, Pic~I, and Pisc~II.

We compute the average SFHs of each kinematic group in two ways. First, we take a simple, unweighted average of the best-fit lookback times as a function of the mass fraction formed.  Second, we compute the weighted average of the same quantities.  For the weighting,  we take the standard deviation of the lookback time at mass fraction $X$, $\sigma(\tau_{X})$, to be half the width of the 68\% confidence interval, and weight by the inverse variance $\sigma({\tau}_{X})^{-2}$.
All reported uncertainties on the averaged SFHs are the standard error on the mean.
We report all average $\tau_{80}$ and $\tau_{90}$ values and uncertainties, and the differences between kinematic groups, in \autoref{tab:t80_by_group}.

In the unweighted case (top panels of \autoref{fig:kinematic_group_sfhs}), we find marginal signal of a delay by comparing population averages.  The main notable differences is a delay between the long-term MW and LMC $\tau_{80}$ in the culled sample ($\Delta \tau_{80} = 1.04 \pm 0.46$) at the level of $\sim2 \sigma$. $\Delta \tau_{90}$ shows no significant difference. A second difference is a modest delay in $\tau_{90}$ of long-term MW satellites vs. the first infall group  in the unculled sample ($\Delta \tau_{80} = -1.02 \pm 0.63$), which is of low statistical significance.  

One challenge in interpreting the unweighted averages is that the sample is dominated by the lowest-mass systems, which have the largest unceratinties in their SFHs.   The amplitude of the SFH uncertainites on these faint systems is larger than the $\sqrt{N}$ gain realized by increasing the sample size.  The result is that the addition of faint galaxies to the unweighted averaging does not result in better constraints on the population mean.  Accordingly, we consider a weighted-averaging scheme to mitigate these effects.

In the weighted scenario, we find larger quenching delays of $\Delta \tau_{80}$~=~1.33-1.65~Gyr between the culled long-term MW and both first-infall and LMC groups, at up to $5\sigma$ significance. 
A major factor for this change is a result of Boo~I dominating the long-term MW group due to its low uncertainties and uniformly ancient SFH.  We also note that the weighted $\tau_{80}$ and $\tau_{90}$ averages for the first-infall and LMC groups are consistently younger than the unweighted values.
The $\Delta \tau_{90}$ values between the long-term MW and LMC groups in the culled sample are the most consistent across the weighted and unweighted averages ($0.80 \pm 0.54$ and $0.83 \pm 0.41$ Gyr respectively). We find a slightly larger delay than that reported by \citet{2021ApJ...920L..19S}, who measured a $\sim$600~Myr delay in $\tau_{90}$ for the LMC group compared to the long-term MW group using the same error-weighted averaging approach. 

While we find a modest signal in the SFHs that could be attributed to patchy reionization, there are several important caveats and considerations.  First, the orbital histories are not well-constrained back to the reionization-era.  We have assumed that the kinematic group traces different ancient environments.  Some simulations suggest that all present-day UFDs were located in comparably low-density regions around the proto-Local Group \citep[e.g.,][]{2019MNRAS.483.4031R}, whereas others expect more environmental variations \citep[e.g.,][]{2018ApJ...856L..22A, 2023ApJ...959...31K}. 
We cannot resolve this tension in the present study and only note it as a caveat. Second, the averaging methods produce different results.  The signal of SFH differences in our weighted average matches or exceeds that reported by \citet{2021ApJ...920L..19S}.  But, results from our unweighted average are more modest. A fair comparision with theory requires that both simulation and observations studies adopt the same population-wide metrics.  Importantly, we have included all data necessary for an interested reader to conduct their own population-level averages.  Finally, some of the variation in quenching timescales is likely driven by processes beyond reionization alone.  The suppression of UFD star formation by reionization is modulated by supernova feedback and subsequent gas accretion, where galaxies that lost their gas supply before reionization could not restart star formation, while those that retained some gas were able to sustain star formation post-reionization \citep[e.g.,][]{2000ApJ...539..517B,2005ApJ...629..259R,2017ApJ...848...85J}. \added{Although the star-forming lifetimes of many of these galaxies are far longer than the timescales on which Type~II SNe occur, several factors play a  role in determining the extent to which a given supernova contributes to the suppression of star formation. These include the location of the supernova with respect to the bulk of the galaxy gas content, the mass of the dark matter halo, and the energetics of the supernova itself \citep[e.g.,][]{2015ApJ...799L..21W, 2015ApJ...807..154B, 2017ApJ...848...85J}.}
These effects, combined with hierarchical assembly processes, may explain the diversity in observed UFD SFHs and highlight the need to account for spatially inhomogeneous reionization when interpreting MW satellite populations.

Despite these caveats, we can place an upper limit on a measureable signal ($\lesssim 800$~Myr) which is consistent with current predictions from simulations of patchy reionization in the local Universe. 
The main obstacle to improving the empirical constraints are the lack of UFDs for which very secure SFHs can be measured and the noise that is introduced by the inherently stochastic SFHs of very sparsely populated systems.

\section{Conclusions} \label{sec:conc}

We have measured deep, uniform \HST photometry and SFHs of 36 UFDs based on CMD fitting.
We recommend observations with $M_{V\rm{eff}} \le -2.5$ ($N^{\star}_{\rm{oMSTO}} \ge 100$) and S/N $\ge$ 100 at the oMSTO for a high-quality SFH fit. Lower S/N ($\sim$50-100) may still be adequate for well-sampled populations ($M_{V\rm{eff}} \lesssim 5$), but robust star-galaxy separation at the oMSTO is important in more marginal cases.

As in prior work, we find moderate evidence of a delay in quenching of star formation in LMC satellites and first-infall galaxies relative to long-term Galactic satellites, although its duration and statistical significance are subject to a number of confounding factors.
We find a global average quenching time of $12.48 \pm 0.18$~Gyr ago, consistent with reionization-driven quenching scenarios, and place an upper limit on the quenching time delay of $\sim800$~Myr, at $2\sigma$ significance. 

In the future, samples of truly isolated field UFDs as well as satellite systems outside the Local Group will provide critical benchmarks for disentangling the effects of patchy reionization vs.\ other environmental and internal factors (e.g., tidal and ram pressure stripping, supernova feedback).

\begin{acknowledgments}

We thank Tom Brown for sharing his star formation history fit results with us.
Support for this work was provided by NASA through grants GO-16293 and AR-17026 from the Space Telescope Science Institute, which is operated by AURA, Inc., under NASA contract NAS5-26555.
The Flatiron Institute is funded by the Simons Foundation. 
This research made use of hips2fits,\footnote{\url{https://alasky.cds.unistra.fr/hips-image-services/hips2fits}} a service provided by CDS.
This work has made use of data from the European Space Agency (ESA) mission \emph{Gaia} (\url{https://www.cosmos.esa.int/gaia}), processed by the \emph{Gaia} Data Processing and Analysis Consortium (DPAC, \url{https://www.cosmos.esa.int/web/gaia/dpac/consortium}). Funding for the DPAC has been provided by national institutions, in particular the institutions participating in the \emph{Gaia} Multilateral Agreement.
This research has made use of NASA’s Astrophysics Data System.
This research has made use of the VizieR catalogue access tool, CDS, Strasbourg, France (DOI: 10.26093/cds/vizier). The original description of the VizieR service was published in \citet{vizier2000}.
This research uses services or data provided by the Astro Data Lab, which is part of the Community Science and Data Center (CSDC) Program of NSF NOIRLab. NOIRLab is operated by the Association of Universities for Research in Astronomy (AURA), Inc. under a cooperative agreement with the U.S. National Science Foundation.
The Pan-STARRS1 Surveys (PS1) and the PS1 public science archive have been made possible through contributions by the Institute for Astronomy, the University of Hawaii, the Pan-STARRS Project Office, the Max-Planck Society and its participating institutes, the Max Planck Institute for Astronomy, Heidelberg and the Max Planck Institute for Extraterrestrial Physics, Garching, The Johns Hopkins University, Durham University, the University of Edinburgh, the Queen's University Belfast, the Harvard-Smithsonian Center for Astrophysics, the Las Cumbres Observatory Global Telescope Network Incorporated, the National Central University of Taiwan, the Space Telescope Science Institute, the National Aeronautics and Space Administration under Grant No. NNX08AR22G issued through the Planetary Science Division of the NASA Science Mission Directorate, the National Science Foundation Grant No. AST-1238877, the University of Maryland, Eotvos Lorand University (ELTE), the Los Alamos National Laboratory, and the Gordon and Betty Moore Foundation.
\end{acknowledgments}

\facilities{HST (ACS).
All the data presented in this paper were obtained from the Mikulski Archive for Space Telescopes (MAST) at the Space Telescope Science Institute. The specific observations analyzed can be accessed via \dataset[DOI: 10.17909/06qe-m148]{https://doi.org/10.17909/06qe-m148}. The HLSP products related to this paper can be accessed via \dataset[DOI: 10.17909/b8yw-wv58]{https://doi.org/10.17909/b8yw-wv58}.
}

\software{Astropy v6.0.1 \citep{2013A&A...558A..33A, 2018AJ....156..123A, 2022ApJ...935..167A}, 
          Astroquery v0.4.7 \citep{2017ascl.soft08004G, 2019AJ....157...98G, 2021zndo....591669G}, 
          DOLPHOT v2.0 \citep{2000PASP..112.1383D, 2016ascl.soft08013D}, 
          Drizzlepac v3.6.1 \citep{2012ascl.soft12011S, 2013ASPC..475...49H, 2015ASPC..495..281A}, 
          \hstonepass v2024.05.29 \citep{2022wfc..rept....5A}, 
          \texttt{MATCH} \citep{2002MNRAS.332...91D}, 
          Matplotlib v3.9.2 \citep{2007CSE.....9...90H, 2021zndo....592536T}, 
          NumPy v1.25.2 \citep{2011CSE....13b..22V, 2020Natur.585..357H}, 
          pandas v2.2.3 \citep{mckinney2010, 2022zndo...3509134T}, 
          seaborn v0.13.2 \citep{2020ascl.soft12015W, 2021zndo....592845W, 2021JOSS....6.3021W}, 
          SciPy v1.10.0 \citep{2020NatMe..17..261V}, 
          Vaex v4.17.0 \citep{2018ascl.soft10004B, 2018A&A...618A..13B}
          }

\appendix

\section{Photometry product descriptions}
\label{app:data}

Here we describe the contents of the DOLPHOT full-stack (source-level) photometry catalogs included in this data release. 
For in-depth descriptions of the DOLPHOT routines and outputs, please see \citet{2000PASP..112.1383D} and the DOLPHOT user manual.\footnote{\url{http://americano.dolphinsim.com/dolphot/dolphot.pdf}}
\autoref{tab:dolphot_cols} describes the columns of the DOLPHOT catalogs, which are provided in FITS tabular format.

\begin{table*}
    \centering
    \caption{Description of columns in the DOLPHOT photometric catalogs.}
    \label{tab:dolphot_cols}
    \begin{tabular}{llll}
    \hline \hline
    Column & Type & Unit & Description\\
    \hline
    ID & str & & Unique source identifier within the field\\
    RA & float32 & deg & Right Ascension (ICRS)\\
    DEC & float32 & deg & Declination (ICRS)\\
    X & float32 & pix & X position in reference image (origin=0.5)\\
    Y & float32 & pix & Y position in reference image (origin=0.5)\\
    OBJTYPE & uint8 & & Object type (1=bright star, 2=faint star)\\
    \hline
    \multicolumn{4}{c}{For each filter$^{a}$}\\
    \hline
    \textit{Filter}\_COUNT & float32 & e$^{-}$ & Total source counts, combined over all images in \textit{Filter}\\
    \textit{Filter}\_SKY & float32 & e$^{-}$ & Total sky level around source, combined over all images in \textit{Filter}\\
    \textit{Filter}\_VEGA & float32 & mag & Vega$^{b}$ magnitude, combined over all images in \textit{Filter}\\
    \textit{Filter}\_ERR & float32 & mag & Uncertainty on  magnitude, combined over all images in \textit{Filter}\\
    \textit{Filter}\_CHI & float32 & & $\chi$ of the PSF fit, combined over all images in \textit{Filter} \\
    \textit{Filter}\_SNR & float32 & & Signal-to-noise ratio of the PSF fit, combined over all images in \textit{Filter} \\
    \textit{Filter}\_SHARP & float32 & & Sharpness of the PSF fit, combined over all images in \textit{Filter} \\
    \textit{Filter}\_ROUND & float32 & & Roundness of the PSF fit, combined over all images in \textit{Filter} \\
    \textit{Filter}\_CROWD & float32 & mag & Crowding of the PSF fit, combined over all images in \textit{Filter}\\
    \textit{Filter}\_FLAG & uint8 & & Quality flag of the PSF fit, combined over all images in \textit{Filter} ($\le 3$ recommended) \\
    \textit{Filter}\_GST & bool & & Does the object pass the GST criteria in this filter? \\
    \hline
    \multicolumn{4}{c}{For each image$^{b}$}\\
    \hline
    \textit{Image}\_COUNT & float32 & e$^{-}$ & Source counts in \textit{Image}\\
    \textit{Image}\_SKY & float32 & e$^{-}$ & Sky level around source in \textit{Image}\\
    \textit{Image}\_VEGA & float32 & mag & Vega magnitude of the PSF fit in \textit{Image}\\
    \textit{Image}\_ERR & float32 & mag & Uncertainty on magnitude of the PSF fit in \textit{Image}\\
    \textit{Image}\_CHI & float32 & & $\chi$ of the PSF fit in \textit{Image}\\
    \textit{Image}\_SNR & float32 & & Signal-to-noise ratio of the PSF fit in \textit{Image}\\
    \textit{Image}\_SHARP & float32 & & Sharpness of the PSF fit in \textit{Image}\\
    \textit{Image}\_ROUND & float32 & & Roundness of the PSF fit in \textit{Image}\\
    \textit{Image}\_CROWD & float32 & mag & Crowding of the PSF fit in \textit{Image}\\
    \textit{Image}\_FLAG & uint8 & & Quality flag of the PSF fit in \textit{Image}\\
    \hline
    \multicolumn{4}{p{0.95\linewidth}}{$^a$ Filters are named as \textit{camera}\_\textit{filter}, for example ACS\_F814W.} \\
    \multicolumn{4}{p{0.95\linewidth}}{$^b$ Zeropoints are based on the CALSPEC Vega spectrum \texttt{alpha\_lyr\_stis\_008.fits}. } \\
    \multicolumn{4}{p{0.95\linewidth}}{$^c$ Images are named as follows: \texttt{ipppssoot}\_\textit{filter}\_\texttt{sfx}\_chip\#, for example jbpb02v9q\_f814w\_flc\_chip1. The \texttt{ipppssoot} and \texttt{sfx} naming conventions are described further in the HST user documentation: \url{https://hst-docs.stsci.edu/hstdhb/2-hst-file-names/2-1-file-name-format}}
    \end{tabular}
\end{table*}

\bibliography{bibliography,software}{}
\bibliographystyle{aasjournal}


\end{document}